\documentclass[10pt,journal]{IEEEtran}

\usepackage{amsmath,amsfonts}
\usepackage{amssymb}   % gives you \square
\usepackage{amsthm}
\usepackage{algorithm}
\usepackage{algorithmic} % (optional: consider algorithmicx+algpseudocode)
\usepackage{array}
\usepackage[caption=false,font=footnotesize,labelfont=sf,textfont=sf]{subfig}
\usepackage{textcomp}
\usepackage{stfloats}
\usepackage{url}
\usepackage{verbatim}
\usepackage{graphicx}
\usepackage{cite}
\usepackage{braket}
\newtheorem{theorem}{Theorem}
\newtheorem{lemma}[theorem]{Lemma}

\newtheorem{definition}[theorem]{Definition}
\newtheorem{corollary}[theorem]{Corollary}

\begin{document}

\title{Entanglement Certification by Measuring Nonlocality}

\author{
Xuan~Du~Trinh, Zhengyu~Wu, Junlin~Bai, Huan\mbox{-}Hsin~Tseng, 
Nengkun~Yu, and Aruna~Balasubramanian%
\thanks{X. D. Trinh, Z. Wu, J. Bai, N. Yu, and A. Balasubramanian, Department of Computer Science, Stony Brook University, Stony Brook, NY 11794, USA (e-mail: \{xtrinh, zhenwu, jubbai, nengkun.yu, arunab\}@cs.stonybrook.edu).}%
\thanks{H.-H. Tseng, AI \& ML Department, Brookhaven National Laboratory, Upton, NY 11973, USA (e-mail: htseng@bnl.gov).}}

% Optional running header (update or remove)
% \markboth{<Journal Name>, Vol.~X, No.~Y, <Month> <Year>}%
% {Trinh \MakeLowercase{\textit{et al.}}: <Short Title>}

% Optional pubid; include only if required by venue
% \IEEEpubid{0000--0000/00\$00.00~\copyright~\the\year\ IEEE}

\maketitle

\begin{abstract}
Reliable verification of entanglement is a central requirement for quantum networks. This paper presents a practical verification approach based on violations of the Clauser–Horne–Shimony–Holt (CHSH) inequality. We derive tight mathematical bounds that relate the CHSH value to entanglement fidelity and introduce a statistical framework that optimizes resource usage while ensuring reliable certification. Our main contributions are: (i) fidelity bounds derived directly from the CHSH measure, which also enable nonlocality certification at sufficiently high fidelities; (ii) a sample-complexity analysis that quantifies the number of measurements required to achieve desired confidence levels for the CHSH measure and the entanglement fidelity; and (iii) verification protocols, some with rigorous mathematical guarantees and others with numerical evaluation. Using NetSquid, we develop a simulation framework that models diverse network conditions and enables systematic exploration of trade-offs in CHSH-based verification. This framework highlights the interplay between accuracy, efficiency, and operational parameters, providing concrete guidelines for deploying entanglement verification in resource-constrained quantum networks.
\end{abstract}


\section{Introduction}

Quantum technologies are demonstrating clear advantages over classical methods in many domains. Quantum algorithms provide major speedups for specific computational problems \cite{Shor1994,Grover1996,Childs2017-solving-linear-system}, quantum machine learning promises improved accuracy and efficiency \cite{Biamonte2017-quantum-ML, Liu2024-quantum-ML, Havlicek2019-quantum-ML}, and quantum cryptography enables provably secure communication \cite{Security-quantum-crypto-Renner}. These advances position quantum networks as a key infrastructure for the future, supporting tasks such as delegated quantum computation, quantum key distribution, and secure long-distance communication \cite{Fitzsimons2017-Blind-QC, Ekert91,Bennett1984-BB84}.

The standard approach for long-distance quantum communication is to generate entangled Einstein–Podolsky–Rosen (EPR) pairs between network nodes and use them as channels for quantum teleportation \cite{Benett-teleportation,Hermans2022-teleportation-networks,Caleffi-quantum-internet}. Intermediate nodes extend communication through entanglement swapping. In practice, however, this process suffers from low success rates. Decoherence, channel loss, and imperfect swapping quickly degrade entanglement, making reliable distribution difficult. Several improvements have been proposed: optimizing resource allocation to extend transmission length \cite{Halder2024}, using parity codes to tolerate photon loss without quantum memories \cite{Munro2012-error-correction}, or applying quantum error correction and purification methods to increase fidelity \cite{Muralidharan2016,Shi2024-routing}. While these results are promising, large-scale deployment still requires efficient and robust verification of entanglement quality.

Existing approaches to entanglement verification include quantum state tomography (QST), which reconstructs the full density matrix of a state \cite{Haah2016-state-tomography,Xin-Yu-state-tomography,HC-NGuyen-State-tomography}. Although reliable, tomography is impractical for networks because of its intensive resource cost. Direct fidelity estimation (DFE) addresses this issue by estimating fidelity relative to a target state with only a small number of Pauli measurement settings, significantly reducing the complexity compared to QST \cite{flammia2011direct,BarberaRodriguez2025}. Entanglement witnesses (EWs) offer another route, requiring fewer measurements to detect entanglement without full reconstruction \cite{Gühne2009-entanglement-detection,Bera2023-entanglement-detection,Siudzinska2021-entanglement-detection,Chruscinski2014-entanglement-detection}.

In this work, we focus on a specific entanglement witness based on the Clauser–Horne–Shimony–Holt (CHSH) inequality, one of the most widely studied Bell inequalities. Unlike many witnesses that only detect the presence of entanglement, CHSH measurements allow us to place both lower and upper bounds on entanglement fidelity concerning the ideal EPR state. This dual capability makes CHSH well suited for network-level verification, offering both theoretical rigor and practical efficiency.

We begin by revisiting the CHSH inequality. We then show how CHSH statistics are connected to entanglement fidelity and present a statistical protocol for fidelity estimation and verification. Finally, we demonstrate the effectiveness of our method in simulated quantum networks using NetSquid \cite{Coopmans2021-NetSquid}, evaluating performance under realistic noise and channel conditions.
\section{Related works and Motivation}

%New version ----------------------

Research on entanglement verification spans several complementary directions, many of which are directly relevant to quantum networks.

% \textit{Quantum teleportation.} Quantum teleportation is a fundamental protocol that enables the transfer of an arbitrary, unknown quantum state from a sender (Alice) to a receiver (Bob) using a shared entangled pair and classical communication \cite{Benett-teleportation}. With a maximally entangled state, the process achieves perfect fidelity through local operations and classical correction. If the shared pair is noisy, teleportation still outperforms any classical scheme, provided the entanglement quality remains above some threshold \cite{horodecki1999general,cavalcanti2017all}. In quantum networks, teleportation enables long-distance communication via entanglement swapping, where intermediate nodes connect shorter links into a larger entangled channel. Since teleportation fidelity depends directly on the quality of the shared pairs, high-confidence entanglement verification becomes essential for reliable network operation.

\textit{Quantum teleportation.} 
Quantum teleportation transfers an arbitrary unknown state from a sender (Alice) to a receiver (Bob) using a shared entangled pair and classical communication \cite{Benett-teleportation}. With maximally entangled pairs, fidelity is perfect. For noisy pairs, teleportation still outperforms any classical scheme if entanglement remains above a threshold \cite{horodecki1999general,cavalcanti2017all}. In quantum networks, teleportation enables long-distance communication via entanglement swapping, which fuses short links into extended channels. Because teleportation fidelity depends directly on pair quality, high-confidence entanglement verification is essential for reliable operations.

\textit{Entanglement purification.} 
First introduced by Bennett et al. \cite{Entanglement-purification-bennett96}, purification protocols employ local operations and classical communication (LOCC) to distill multiple noisy entangled pairs into fewer high-fidelity ones. This improves the effectiveness of verification by supplying higher-quality entanglement \cite{PhysRevResearch.5.033171-entanglement-purification,PhysRevLett.128.080504-entanglement-purification,Yan2023-entanglement-purification}.

\textit{Direct Fidelity Estimation.} DFE offers a scalable alternative to full QST for evaluating the fidelity of a prepared state against a known pure target \cite{flammia2011direct,BarberaRodriguez2025}. Instead of reconstructing the full density matrix, DFE samples only a tailored subset of Pauli operators. Crucially, the number of required measurement settings depends only on the desired statistical accuracy, not on the Hilbert space dimension. For a system of dimension $d$, DFE needs about $d$ times fewer Pauli measurement settings than QST, which typically requires $d^2$. This advantage makes DFE especially suitable for application in multi-qubit systems.

% \textit{Existing GHZ state verification protocols.}  
% Well-established techniques exist for verifying multipartite entangled states such as GHZ. McCutcheon \textit{et al.} demonstrated a protocol that allows any network node to verify genuine multipartite entanglement under realistic conditions, including untrusted parties and imperfect devices \cite{McCutcheon2016-GHZ-verification}. Their approach uses physically accessible local measurements to certify closeness to a target state. Importantly, the same framework can also be applied to the bipartite case, where the EPR pair corresponds to the $n=2$ instance of a GHZ state. However, these protocols typically provide only one-sided guarantees (e.g., lower bounds) on fidelity. While effective, this limitation highlights the need for complementary methods that offer stronger certification tailored to EPR pairs. Since high-fidelity EPR entanglement forms the foundation for building larger network states, reliable bipartite verification is a critical first step toward scalable quantum network architectures.

\textit{Existing GHZ state verification protocols.}  
Well-established techniques exist for verifying multipartite entangled states such as GHZ. McCutcheon \textit{et al.} demonstrated a protocol that allows to verify genuine multipartite entanglement under realistic conditions, including untrusted parties and imperfect devices \cite{McCutcheon2016-GHZ-verification}. The method relies on accessible local measurements to certify closeness to a target state. Importantly, the same framework can also be restricted to the bipartite case, where the EPR pair corresponds to the $n=2$ instance of a GHZ state. However, the protocol typically provides only one-sided guarantees (e.g., lower bounds) on fidelity. This limitation motivates complementary approaches that deliver stronger certification for EPR pairs, the most fundamental building block of distributed quantum information processing, from which larger structures such as GHZ states and scalable quantum networks can be constructed.

\textit{Entanglement witnesses.} EWs provide a well-established framework for detecting entanglement in multipartite systems. An EW is an observable whose expectation value offers a sufficient condition for entanglement. A major advantage is resource efficiency: carefully designed witnesses require only moderate measurement overhead, balancing verification accuracy with experimental cost \cite{Gühne2009-entanglement-detection,Bera2023-entanglement-detection,Siudzinska2021-entanglement-detection,Chruscinski2014-entanglement-detection}. In distributed settings, EWs can be evaluated via LOCC, where each node measures locally and shares outcomes to compute the witness value \cite{PhysRevA.66.062305-entanglement-detection-few-local-measurement}. If the value exceeds a given threshold, entanglement is certified. More recently, machine learning methods have been used to enhance robustness against noise and incomplete data, opening new possibilities for scalable certification in realistic networks \cite{Asif2023-EntanglementDetection2023-deep-learning,Urena-EntanglementDetection2024-deep-learning,PhysRevApplied.19.034058-entanglement-detection-machine-learning}.

\textit{CHSH measure – Our motivation.}  
While entanglement witnesses are valuable, they often only detect whether entanglement exists without quantifying its strength. For our purposes, we require a witness directly tied to entanglement fidelity, enabling both detection and quantitative assessment. The CHSH inequality provides exactly this: its violation can be linked to fidelity with an ideal EPR pair, making it well suited for bipartite certification in quantum networks.  

Our choice of CHSH-based verification is motivated by three considerations. First, quantum nonlocality remains central to quantum information theory, connecting foundational physics with advanced applications. Second, CHSH violation underpins device-independent quantum key distribution (DIQKD), where Bell inequality violations directly certify security \cite{Zapatero2023-DI-QKD}. This dual role—as both fidelity benchmark and device-independent primitive—makes CHSH a unifying tool for networks supporting multiple protocols. Third, CHSH is the most studied Bell inequality, with demonstrated feasibility: loophole-free violations have been achieved over kilometer-scale channels \cite{hensen2015loophole,giustina2015significant,shalm2015strong-loophole-free-test}, and recent experiments applied CHSH-based DIQKD over hundreds of meters \cite{zhang2022device}. These results confirm that CHSH-based verification is practical under realistic conditions, underscoring its relevance for near-term quantum networks.
\section{Problem Statement and Preliminaries}

% Our work focuses on the problem of certifying high-quality entanglement generated between two nodes of a quantum network, operated by Alice and Bob. We consider a scenario in which a device distributes EPR pairs to the quantum memories of Alice and Bob for the purpose of teleportation. In the ideal case, each shared pair is in the Bell state $\ket{\Phi^+} = \frac{\ket{00} + \ket{11}}{\sqrt{2}}.$
% As noise is unavoidable, the actual state of each EPR pair should be represented by an element in the set of all four-dimensional density operators, denoted by $\mathcal{Q}$. Thanks to the comprehensive capacity to represent probabilistic mixture among any ensemble of quantum states, we assume that the device consistently generates identical and independently distributed (i.i.d) noisy EPR states $\rho \in \mathcal{Q}$. A classical communication channel between Alice and Bob is also available for exchanging measurement settings and outcomes during the verification procedure. Quantitatively, our goal is to develop methods for estimating and statistically testing the entanglement fidelity of $\rho$ with respect to $\ket{\Phi^+}$.

Our work addresses the certification of high-quality entanglement between two network nodes, Alice and Bob. A source distributes EPR pairs to their quantum memories for teleportation. In the ideal case, each shared pair is in the Bell state $\ket{\Phi^+} = \frac{\ket{00} + \ket{11}}{\sqrt{2}}.$ As noise is unavoidable, the actual state of each EPR pair should be represented by an element in the set of all four-dimensional density operators, denoted by $\mathcal{Q}$. We assume that the device consistently generates identical and independently distributed (i.i.d) noisy EPR states $\rho \in \mathcal{Q}$. A classical channel allows Alice and Bob to exchange measurement settings and outcomes. Our goal is to design methods for estimating and statistically testing the entanglement fidelity of $\rho$ relative to $\ket{\Phi^+}$.

\begin{definition}
    The entanglement fidelity of quantum state $\rho\in \mathcal{Q}$ is defined by $F(\rho)=\bra{\Phi^+}\rho\ket{\Phi^+}$.  
\end{definition}
\noindent
\textbf{Remark.} The unique state that achieves the maximal entanglement fidelity $1$ is $\rho = \ket{\Phi^+}\bra{\Phi^+}$. Indeed, by Cauchy-Schwartz inequality, we have
    \begin{align}
        F(\rho)=\text{Tr}[\rho\ket{\Phi^+}\bra{\Phi^+}]\leq \sqrt{\text{Tr}[\rho^2]}\leq 1.
    \end{align}
If $F(\rho)=1$, then $\text{Tr}[\rho^2] = 1$ and $\text{rank}(\rho)=1$. That implies $\rho = \ket{\phi}\bra{\phi}$ with $|\braket{\phi|\Phi^+}|=1.$\qed

\begin{figure}
    \centering
    \includegraphics[width=0.8\linewidth]{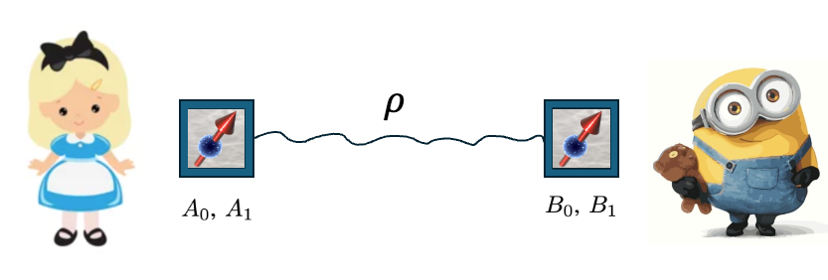}
    \caption{Alice and Bob share noisy state $\rho$. In each measurement round, Alice chooses one observable between $A_0$ and $A_1$, and similarly Bob picks one observable between $B_0$ and $B_1$.}
    \label{fig: experiment setup}
\end{figure}

% \begin{definition}
% The \textit{CHSH measure} is based on a Bell-type experiment where two parties, \textbf{Alice} and \textbf{Bob}, perform independent measurements on a bipartite quantum state $\rho\in \mathcal{Q}$. Each party chooses one of two measurement settings:
% \begin{itemize}
%     \item Alice chooses between observables \( A_0 \) and \( A_1 \).
%     \item Bob chooses between observables \( B_0 \) and \( B_1 \).
% \end{itemize}
% Each measurement has binary outcomes, conventionally labeled as \( \pm1 \). The CHSH measure \( S \) is defined as:
% \begin{equation}
%     S = \langle A_0 B_0 \rangle + \langle A_1 B_0 \rangle + \langle A_0 B_1 \rangle - \langle A_1 B_1 \rangle.
% \end{equation}
% where \( \langle A_i B_j \rangle \) represents the expectation value of the product of measurement outcomes when Alice uses setting \( A_i \) and Bob uses setting \( B_j \) (see Figure \ref{fig: experiment setup}).    
% \end{definition}

\begin{definition}
The \textit{CHSH measure} is defined for a Bell-type experiment in which two parties, Alice and Bob, perform local measurements on a bipartite quantum state $\rho\in\mathcal{Q}$. Each party chooses one of two settings:
\begin{itemize}
    \item Alice chooses between observables \( A_0 \) and \( A_1 \).
    \item Bob chooses between observables \( B_0 \) and \( B_1 \).
\end{itemize}
with binary outcomes in $\{\pm1\}$. The CHSH value \(S\) is
\begin{equation}
    S \;=\; \langle A_0 B_0 \rangle + \langle A_1 B_0 \rangle + \langle A_0 B_1 \rangle - \langle A_1 B_1 \rangle,
\end{equation}
where $\langle A_i B_j\rangle$ denotes the expectation of the product of the outcomes when Alice uses $A_i$ and Bob uses $B_j$ (see Fig.~\ref{fig: experiment setup}). 
% Optionally, make the correlator explicit:
% $\langle A_i B_j\rangle \equiv \operatorname{Tr}\!\big[\rho\,(A_i\!\otimes\! B_j)\big]$.
\end{definition}

In general, as a result induced from \cite{PhysRevLett.23.880-CHSH-inequality}, in any local hidden variable theory, the absolute value of \( S \) satisfies the CHSH inequality:
\begin{equation}
    |S| \leq 2.
    \label{Bell local hidden variable inequality}
\end{equation}
However, the local hidden variable model is pointed out to be incompatible with quantum mechanics. The inequality \eqref{Bell local hidden variable inequality} is violated with $S = 2\sqrt{2}$ in the well-known quantum setting where $\rho=\ket{\Phi^+}\bra{\Phi^+}$  and
\begin{align}
    A_0 &= \sigma_x=\begin{bmatrix} 0 & 1 \\ 1 & 0 \end{bmatrix}, \quad &&A_1 = \sigma_z=\begin{bmatrix} 1 & 0 \\ 0 & -1 \end{bmatrix}, \notag\\
    B_0 &= \frac{\sigma_x + \sigma_z}{\sqrt{2}}, &&B_1 = \frac{\sigma_x - \sigma_z}{\sqrt{2}}.
    \label{Eq: Choice of observables}
\end{align}
Additionally, Tsirelson showed that $2\sqrt{2}$ is also the maximal value of $|S|$ that a quantum setting can achieve \cite{Cirelson1980}. Formally, we have the following result.

\begin{theorem}[Tsirelson] 
    {Define an} operator version of the CHSH measure as
    \begin{equation}
         \hat{S} = A_0 \otimes B_0 + A_1 \otimes B_0 + A_0 \otimes B_1 - A_1 \otimes B_1.
    \end{equation}
    For any $\rho\in \mathcal{Q}$ and any choice of observables $A_0,A_1,B_0,B_1$ with outcome $\pm 1$, the following inequality holds
    \begin{equation}
    |S| = |\text{Tr} [\hat{S}\rho ] | \leq 2\sqrt{2}.
\end{equation}
\end{theorem}

\noindent
\textbf{Remark.} Before presenting the main results, we note:

\begin{itemize}
    \item $\mathcal{Q}$ is a convex subset of $\mathbf{M}(\mathbb{C},4\times 4)$. The map $S:\mathcal{Q}\to\mathbb{R}$, $\rho\mapsto \mathrm{Tr}[\rho \hat{S}]$, is linear (hence convex). Since $\mathcal{Q}$ is closed and bounded (see Supplementary Material A), it is compact by the Heine–Borel theorem, and $S$ attains its global maximum and minimum.
    \item The Bell basis of maximally entangled states is
    \begin{align}
        |\Phi^\pm\rangle &= \tfrac{|00\rangle \pm |11\rangle}{\sqrt{2}}, \quad
        |\Psi^\pm\rangle = \tfrac{|01\rangle \pm |10\rangle}{\sqrt{2}},
        \label{Eq: Bell basis}
    \end{align}
    and for the observables in Eq.~\eqref{Eq: Choice of observables}, the CHSH operator admits the Bell-diagonal form
    \begin{equation}
        \hat{S} = 2\sqrt{2}\left(|\Phi^+\rangle\langle\Phi^+| - |\Psi^-\rangle\langle\Psi^-|\right).
        \label{E: Bell-diagonal form}
    \end{equation}
\end{itemize}

% This implies a global maximum $S(|\Phi^+\rangle\langle\Phi^+|)=2\sqrt{2}$ and a global minimum $S(|\Psi^-\rangle\langle\Psi^-|) = -2\sqrt{2}$.
% \end{itemize}

% \begin{align}
%     \left( 2\sqrt{2}, |\Phi^+\rangle \right), \left( 0, |\Phi^-\rangle \right), \left( 0, |\Psi^+\rangle \right),    \left( -2\sqrt{2}, |\Psi^-\rangle \right).
% \end{align}

\section{Main Results}

% \subsection{Explicit experiment setting}
% \begin{align}
%     A_0 &= \sigma_x=\begin{bmatrix} 0 & 1 \\ 1 & 0 \end{bmatrix}, \quad A_1 = \sigma_z=\begin{bmatrix} 1 & 0 \\ 0 & -1 \end{bmatrix}, \\
%     B_0 &= \frac{\sigma_x + \sigma_z}{\sqrt{2}}=\frac{1}{\sqrt{2}} \begin{bmatrix} 1 & 1 \\ 1 & -1 \end{bmatrix}, \quad B_1 = \frac{\sigma_x - \sigma_z}{\sqrt{2}}=\frac{1}{\sqrt{2}} \begin{bmatrix} -1 & 1 \\ 1 & 1 \end{bmatrix}.
% \end{align}

% The CHSH operator is defined as:
% \begin{equation}
%     \hat{S} = A_0 \otimes B_0 + A_1 \otimes B_0 + A_0 \otimes B_1 - A_1 \otimes B_1.
% \end{equation}
% The explicit matrix form is:

% \begin{equation}
% \hat{S} =
% \begin{bmatrix}
% \sqrt{2} & 0 & 0 & \sqrt{2} \\
% 0 & -\sqrt{2} & \sqrt{2} & 0 \\
% 0 & \sqrt{2} & -\sqrt{2} & 0 \\
% \sqrt{2} & 0 & 0 & \sqrt{2}
% \end{bmatrix}.
% \end{equation}

% \begin{align}
%     |\Phi^+\rangle &= \begin{bmatrix} 0.7071 \\ 0 \\ 0 \\ 0.7071 \end{bmatrix}, \\
%     |\Psi^+\rangle &= \begin{bmatrix} 0 \\ 0.7071 \\ 0.7071 \\ 0 \end{bmatrix}, \\
%     |\Psi^-\rangle &= \begin{bmatrix} 0 \\ 0.7071 \\ -0.7071 \\ 0 \end{bmatrix}, \\
%     |\Phi^-\rangle &= \begin{bmatrix} 0.7071 \\ 0 \\ 0 \\ -0.7071 \end{bmatrix}.
% \end{align}

\textbf{From now, we explicitly fix the choice of observables as in Equation \eqref{Eq: Choice of observables} for the experiment setting between Alice and Bob. Then, $\hat{S}$ is explicitly fixed as in Equation (\ref{E: Bell-diagonal form}).}
\subsection{\textbf{Bounds for entanglement fidelity}}
The following theorem provides a tool to bound the entanglement fidelity of an arbitrary two-qubit quantum state.
\begin{theorem}
For any quantum state \( \rho \in \mathcal{Q} \), the corresponding entanglement fidelity satisfies the following inequalities:
\begin{equation}
    \frac{S(\rho)}{2\sqrt{2}} \leq F(\rho) \leq \frac{S(\rho)}{4\sqrt{2}} + \frac{1}{2}.
    \label{eq: theorem bounds}
\end{equation}
\label{theorem: bound entanglement fidelity}
\end{theorem}

\textit{Proof.} We prove the lower bound using the definition of entanglement fidelity. Let \( \rho \in \mathcal{Q} \), then
\begin{align}
    S(\rho) &= 2\sqrt{2} \big[ \bra{\Phi^+} \rho \ket{\Phi^+} - \bra{\Psi^-} \rho \ket{\Psi^-} \big] \notag\\
    &\leq 2\sqrt{2} \bra{\Phi^+} \rho \ket{\Phi^+} \quad (\text{since } \rho \geq 0)\notag\\
    &= 2\sqrt{2} F(\rho).
\end{align}

The upper bound follows from the unit trace of $\rho$. Since $\rho$ is positive semidefinite with $\mathrm{Tr}(\rho)=1$, all diagonal elements are nonnegative and sum to one:
\begin{equation}
    \bra{\Phi^+}\rho\ket{\Phi^+} +
    \bra{\Phi^-}\rho\ket{\Phi^-} +
    \bra{\Psi^+}\rho\ket{\Psi^+} +
    \bra{\Psi^-}\rho\ket{\Psi^-} = 1.
\end{equation}
This implies that $\bra{\Phi^+}\rho\ket{\Phi^+}+\bra{\Psi^-}\rho\ket{\Psi^-}\leq 1$. Combining this inequality with the definition of $S(\rho)$, we deduce that
\begin{align}
    S(\rho) &= 2\sqrt{2}[\bra{\Phi^+}\rho\ket{\Phi^+}-\bra{\Psi^-}\rho\ket{\Psi^-}]\notag\\
    &\geq 2\sqrt{2}\left[2\bra{\Phi^+}\rho\ket{\Phi^+}-1\right]\notag\\
    &=2\sqrt{2}\left[2F(\rho)-1\right].
\end{align}
Rearranging the inequality gives $F(\rho) \leq \frac{S(\rho)}{4\sqrt{2}} + \frac{1}{2}. \qed$

\begin{corollary}
The EPR state $\rho = \ket{\Phi^+}\bra{\Phi^+}$ is the unique global maximum of $S$ over $\mathcal{Q}$.
\label{Cor: unique global maximum}
\end{corollary}
\textit{Proof.} Let $\rho \in \mathcal{Q}$ that achieves the maximum $S(\rho)=2\sqrt{2}$. So, $1\geq F(\rho)\geq \frac{S(\rho)}{2\sqrt{2}}=1\Rightarrow F(\rho) =1$. Because $\rho$ achieves the maximal entanglement fidelity, $\rho = \ket{\Phi^+}\bra{\Phi^+}$ as remarked above. \qed

% \begin{corollary}
% Assume that after the preparation, the EPR state is affected by a noisy channel and the actual noisy state $\rho$ gives $S(\rho)=2\sqrt{2}-\epsilon$ for some small $0<\epsilon <2\sqrt{2}$, then we have that $F(\rho) \geq 1 - \frac{\epsilon}{2\sqrt{2}}.$
% \end{corollary}

\subsection{\textbf{Nonlocality certification via entanglement fidelity}}

Before analyzing the cost of quantitative entanglement certification via nonlocality measurements, we first consider the inverse task: \textit{certifying nonlocality from entanglement fidelity} using Theorem~\ref{theorem: bound entanglement fidelity}.

A natural question is: for which values of fidelity $F(\rho)$ can we guarantee that a state $\rho$ is entangled or violates the CHSH inequality? For intuition, consider the example of Werner's state
\begin{equation}
    \rho = \frac{4F(\rho)-1}{3}\ket{\Phi^+}\bra{\Phi^+} 
         + \frac{4(1-F(\rho))}{3}\frac{I}{2}\otimes \frac{I}{2}.
\end{equation}
This state is entangled for $F(\rho) > \tfrac{1}{2}$ and violates CHSH when $F(\rho) > \tfrac{3+\sqrt{2}}{4\sqrt{2}} \approx 0.7803$.

For a general state $\rho \in \mathcal{Q}$, Theorem~\ref{theorem: bound entanglement fidelity} gives the sufficient condition $F(\rho) > \tfrac{1+\sqrt{2}}{2\sqrt{2}} \approx 0.8536$
for CHSH violation (and hence entanglement), with fidelity taken with respect to $\ket{\Phi^+}$. Since any maximally entangled two-qubit pure state is locally equivalent to $\ket{\Phi^+}$, this result extends to arbitrary maximally entangled references.

\begin{corollary}[Nonlocality certification from fidelity]
If $\rho \in \mathcal{Q}$ satisfies
\[
    \bra{\phi}\rho\ket{\phi} > \tfrac{1+\sqrt{2}}{2\sqrt{2}}
\]
for some maximally entangled $\ket{\phi}$, then $\rho$ violates the CHSH inequality.
\end{corollary}

\textit{Proof.} Let $U_L$ be a local unitary with $U_L\ket{\Phi^+}=\ket{\phi}$. Then $U_L^\dagger \hat{S} U_L$ is a valid CHSH operator, and applying Theorem~\ref{theorem: bound entanglement fidelity} to $U_L\rho U_L^\dagger$ yields
\begin{equation}
    \text{Tr}[(U_L^\dagger \hat{S} U_L)\rho]
        = S(U_L\rho U_L^\dagger)\geq 4\sqrt{2}\!\left(\bra{\phi}\rho\ket{\phi}-\tfrac{1}{2}\right)> 2.
\end{equation}
Thus $\rho$ exceeds the classical CHSH bound and is nonlocal. \qed

\subsection{\textbf{Statistical Estimation for CHSH}}

From Theorem~\ref{theorem: bound entanglement fidelity}, if $S$ approaches the maximal quantum violation $2\sqrt{2}$, then the underlying state is close to the EPR pair. Motivated by this, we develop statistical methods to test entanglement quality. The fidelity bounds tighten as $S$ increases, with the gap between upper and lower bounds given by $\tfrac{1}{2}-\tfrac{S(\rho)}{4\sqrt{2}}$. We estimate $S$ using Algorithm~\ref{Alg: measure CHSH} and analyze statistical errors.
\begin{algorithm}[h!]
\caption{Estimation of CHSH Measure}
\begin{algorithmic}[1]
\STATE \textbf{Input:} Number of samples per setting $N$ (total measurements $4N$).
\STATE Prepare $4N$ copies of $\rho$.
\FOR{each basis pair $(A_i,B_j)$, $i,j\in\{0,1\}$}
    \STATE Measure $N$ copies, record outcomes $(a_k,b_k)$ for $k=1,\dots,N$.
    \STATE Estimate correlation:
    \[
       \overline{A_iB_j} = \tfrac{1}{N}\sum_{k=1}^N a_k b_k.
    \]
\ENDFOR
% \STATE Compute CHSH value:
% \[
%    \Bar{S} = \overline{A_0B_0} + \overline{A_1B_0} + \overline{A_0B_1} - \overline{A_1B_1}.
% \]
\STATE \textbf{Output:} $\Bar{S}=\overline{A_0B_0} + \overline{A_1B_0} + \overline{A_0B_1} - \overline{A_1B_1}$.
\end{algorithmic}
\label{Alg: measure CHSH}
\end{algorithm}

% Statistically, each correlation term \( \langle A_i B_j \rangle = \mathbb{E}[ A_i B_j ] \) is estimated from a finite number of measurements. In expectation, 
% \begin{equation}
%     \mathbb{E}[ A_i B_j ] = \mathbb{E}[\overline{A_i B_j}]= \frac{1}{N} \sum_{k=1}^{N} \mathbb{E}[ a_k b_k ],
% \end{equation}
% where the product \( a_kb_k \in \{-1, +1\} \) are measurement outcomes, and \( N \) is the number of trials for each setting, meaning that $\mathbb{E}[\bar{S}]=S(\rho)$. The variance of each correlation term follows:
% \begin{equation}
%     \text{Var}[\overline{ A_i B_j} ] = \frac{1 - \mathbb{E}[ A_i B_j]^2}{N}.
% \end{equation}
% By error propagation, the total variance in \( \Bar{S} \) is:
% \begin{align}
% \text{Var}[\Bar{S}] &=\sum_{i,j}\text{Var}[A_i B_j ] \notag\\
%     &=\sum_{i,j} \frac{1 - \mathbb{E}[ A_i B_j]^2}{N}\notag\\
%     &=\frac{4-\sum_{i,j}\mathbb{E}[ A_i B_j]^2}{N}\notag\\
%     &\leq \frac{4-\frac{1}{4}[S(\rho)]^2}{N} \quad (\text{Cauchy-Schwarz}).
% \end{align}

% At this point, we need to have an idea about how many copies of $\rho$ that is sufficient to estimate the CHSH measure with high precision. Below, Lemma \ref{lemma: Using Chebyshev} and Lemma \ref{lemma: Using Hoeffding} provides two formulas for this purpose, and Theorem \ref{theorem: number of copies for CHSH} is their corollary.

Statistically, each correlation term $\langle A_i B_j \rangle = \mathbb{E}[A_iB_j]$ is estimated from $N$ independent trials. The estimator
\begin{equation}
    \overline{A_iB_j} = \frac{1}{N}\sum_{k=1}^N a_k b_k
\end{equation}
satisfies $\mathbb{E}[\overline{A_iB_j}] = \langle A_iB_j \rangle$ with $ a_k b_k \in \{\pm 1\}$, so that $\mathbb{E}[\Bar{S}] = S(\rho)$. The variance of each correlation term
follows:
\begin{equation}
    \text{Var}[\overline{A_iB_j}] = \frac{1-\langle A_iB_j \rangle^2}{N}.
\end{equation}
By error propagation and {Cauchy-Schwarz} inequality,
\begin{align}
    \text{Var}[\Bar{S}] = \frac{4 - \sum_{i,j}\langle A_iB_j \rangle^2}{N} \leq \frac{4 - \tfrac{1}{4}[S(\rho)]^2}{N} .
\end{align}

Thus, the variance decreases as $1/N$, and concentration inequalities can be used to bound the number of copies of $\rho$ required for high-precision CHSH estimation. In what follows, Lemma~\ref{lemma: Using Chebyshev} and Lemma~\ref{lemma: Using Hoeffding} provide two such bounds, with Theorem~\ref{theorem: number of copies for CHSH} as their corollary.
\begin{lemma}
For $\epsilon \in (0, 2\sqrt{2}) $ and $\delta \in (0,1)$, to obtain an estimate $\Bar{S}$ of $S(\rho)$ within $(S(\rho)-\epsilon,\,S(\rho)+\epsilon)$ with probability at least $1-\delta$, it suffices to use
 $N=O(\frac{1}{\delta\epsilon^2})$.
    \label{lemma: Using Chebyshev}
\end{lemma}
\textit{Proof.} By Chebyshev's inequality, we have the following chain of inequalities:
\begin{equation}
\mathbb{P}[|\Bar{S}-S(\rho)|>\epsilon]\leq\frac{\text{Var}[\Bar{S}]}{\epsilon^2} \leq \frac{4-\frac{1}{4}[S(\rho)]^2}{N\epsilon^2} \leq \frac{4}{N\epsilon^2}.
\end{equation}

Then, the sufficient number of copies so that $\mathbb{P}[|\Bar{S}-S(\rho)|>\epsilon]\leq \delta$ is $N=\frac{4}{\delta\epsilon^2}$. If $\rho$ is nonlocal, $S(\rho) \geq 2$, then we need $N=\frac{3}{\delta\epsilon^2}.\qed$

\begin{lemma}
    For $\epsilon \in (0, 2\sqrt{2}) $ and $\delta \in (0,1)$, to obtain an estimate $\Bar{S}$ of $S(\rho)$ within $(S(\rho)-\epsilon,\,S(\rho)+\epsilon)$ with probability at least $1-\delta$, it suffices to use $N=O(\frac{1}{\epsilon^2}\ln{\frac{2}{1-(1-\delta)^\frac{1}{4}}})$.
    \label{lemma: Using Hoeffding}
\end{lemma}
% \textit{Proof.} We apply Hoeffding's inequality for probability distribution of $\overline{A_iB_j}$ over range of values $[-1,+1]$ to obtain $\forall i,j \in \{0,1\}$,
% \begin{align}
%     \mathbb{P}\left[|\overline{A_iB_j} - \mathbb{E}[A_iB_j]|\leq\frac{\epsilon}{4}\right]&\geq 1 - 2\exp{\left(-\frac{N\epsilon^2}{32}\right)}  .
% \end{align}
% Consider the two events
% \begin{equation}
%     E_1 =\left\{|\bar{S}-S(\rho)|\leq \epsilon\right\},
% \end{equation}
% \begin{equation}
%      E_2 =\left\{\bigcap_{i,j\in \{0,1\}}\left(|\overline{A_iB_j} - \mathbb{E}[A_iB_j]|\leq \frac{\epsilon}{4}\right)\right\}.
% \end{equation}
% Note that event $E_2$ implies event $E_1$. So, we have $\mathbb{P}\left[E_1\right]\geq \mathbb{P}\left[E_2\right]$. Thus,
% \begin{align}
%     \mathbb{P}\left[|\bar{S}-S(\rho)|\leq \epsilon\right]
%     &\geq \left(1 - 2\exp{\left(-\frac{N\epsilon^2}{32}\right)}\right)^4
% \end{align}
% So, to ensure $\mathbb{P}\left[|\bar{S}-S(\rho)|> \epsilon\right]\leq\delta$, the input $N$ needs to satisfy $N\geq    \frac{32}{\epsilon^2}\ln{\frac{2}{1-(1-\delta)^\frac{1}{4}}}.$\qed

\textit{Proof.} By Hoeffding’s inequality applied to $\overline{A_iB_j}$ with range $[-1,1]$, for all $i,j\in\{0,1\}$ we have
\begin{align}
    \mathbb{P}\!\left[\,|\overline{A_iB_j}-\mathbb{E}[A_iB_j]|\leq \tfrac{\epsilon}{4}\,\right]
        \geq 1-2\exp\!\left(-\tfrac{N\epsilon^2}{32}\right).
\end{align}
Define the events
\begin{align}
    E_1 &= \{\,|\Bar{S}-S(\rho)|\leq \epsilon\,\}, \\
    E_2 &= \bigcap_{i,j\in\{0,1\}}\{\,|\overline{A_iB_j}-\mathbb{E}[A_iB_j]|\leq \tfrac{\epsilon}{4}\,\}.
\end{align}
Since $E_2 \subseteq E_1$, it follows that $\mathbb{P}[E_1]\geq \mathbb{P}[E_2]$. Therefore,
\begin{align}
    \mathbb{P}[|\Bar{S}-S(\rho)|\leq \epsilon]
        \geq \left(1-2\exp\!\left(-\tfrac{N\epsilon^2}{32}\right)\right)^4.
\end{align}
To ensure $\mathbb{P}[|\Bar{S}-S(\rho)|> \epsilon]\leq \delta$, it suffices that
\begin{equation}
    N \;\geq\; \tfrac{32}{\epsilon^2}\ln\!\Biggl(\tfrac{2}{1-(1-\delta)^{1/4}}\Biggr). \quad \qed
\end{equation}

% Between the two sufficient numbers of copies obtained in Lemma \ref{lemma: Using Chebyshev} and Lemma \ref{lemma: Using Hoeffding}, which one is smaller, we can use it to ensure the high confidence interval of our CHSH estimation. While fixing $\epsilon =0.05$, we present the number of copies needed for $\delta$ ranging from $0.001$ to $0.05$ in Fig.~\ref{fig: sufficient number of copies}. When we require very high confidence in our estimation of $S(\rho)$, with $\delta < \delta_{\text{threshold}} \approx 0.01493$, the upper bound given by Hoeffding's inequality is better. Otherwise, we use Chebyshev's bound. Definitely, this result is at a theoretical level to ensure a certain level of precision. In practice, the necessary and sufficient number could be much more economical since the two inequalities are not totally tight. We conclude the analysis for CHSH estimation with Theorem \ref{theorem: number of copies for CHSH}.

\begin{figure}
    \centering
    \includegraphics[width=0.6\linewidth]{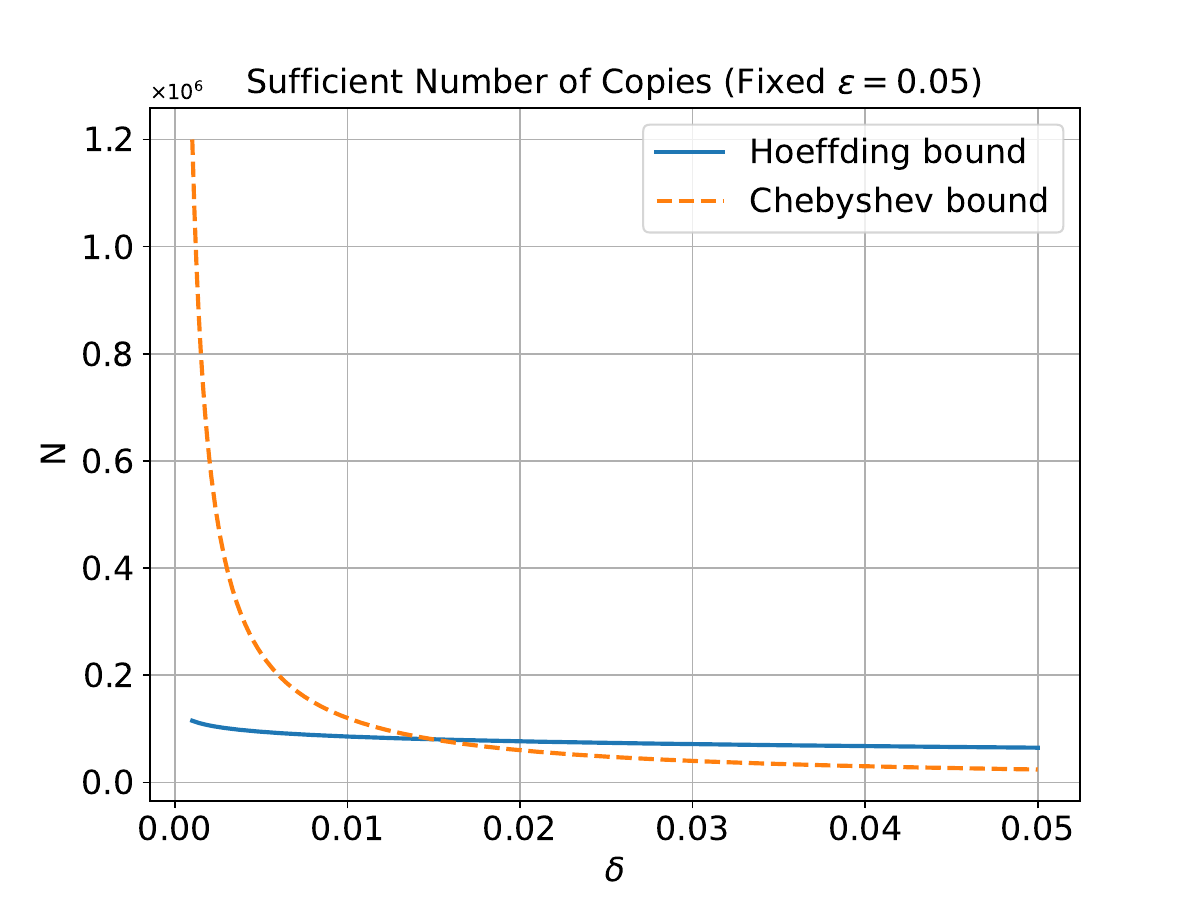}
    \caption{Sufficient number of copies to ensure high-fidelity (probability of $1-\delta$) estimation range of $S(\rho)$, with $\epsilon$ fixed at 0.05}
    \label{fig: sufficient number of copies}
\end{figure}

Between the two sufficient sample sizes from Lemma~\ref{lemma: Using Chebyshev} and Lemma~\ref{lemma: Using Hoeffding}, we take the minimum to guarantee an $\epsilon$–accurate estimate $\Bar{S}$ with confidence $1-\delta$. Fixing $\epsilon=0.05$, Fig.~\ref{fig: sufficient number of copies} plots the required number of copies versus $\delta\in[0.001,0.05]$. For very high confidence ($\delta<\delta_{\text{thr}}\approx 0.01493$), the Hoeffding bound is tighter; otherwise, Chebyshev’s bound dominates. These are sufficient bounds and practical requirements are typically smaller since the inequalities are not absolutely tight. We conclude the CHSH–estimation analysis with Theorem~\ref{theorem: number of copies for CHSH}.

\begin{theorem}
        For $\epsilon \in (0, 2\sqrt{2}) $ and $\delta \in (0,1)$, to obtain an estimate $\Bar{S}$ of $S(\rho)$ within $(S(\rho)-\epsilon,\,S(\rho)+\epsilon)$ with probability at least $1-\delta$, it suffices to use $N=\min\{\frac{3}{\delta\epsilon^2},\frac{32}{\epsilon^2}\ln{\frac{2}{1-(1-\delta)^\frac{1}{4}}}\}$.
    \label{theorem: number of copies for CHSH}
\end{theorem}

\begin{corollary}
   It suffices to run Algorithm \ref{Alg: measure CHSH} with input $N = \min \{\frac{3}{\delta\epsilon^2}, \frac{32}{\epsilon^2}\ln{\frac{2}{1-(1-\delta)^\frac{1}{4}}}\}$ to bound the entanglement fidelity of $\rho$ in the range $\left[\frac{\bar{S}}{2\sqrt{2}}-\frac{\epsilon}{2\sqrt{2}},\min\{\frac{\bar{S}}{4\sqrt{2}}+\frac{1}{2}+\frac{\epsilon}{4\sqrt{2}},1\}\right]$ with probability at least $1-\delta$.
    \label{Cor: Estimate Fidelity using chsh}
\end{corollary}

% \textit{Example.} Set $\delta=0.01$, $\epsilon=0.05$, it suffices to use $N=85.5 \times 10^3$ to get a confidence interval with probability $0.99$ (by Hoeffding). If we tolerate a higher error probability at $\delta=0.05$, the sufficient input is only $N=24\times 10^3$ (by Chebyshev). If the state is measured with $\bar{S}=2\sqrt{2}-0.03$, the high confidence interval for the entanglement fidelity is $[0.972,1]$.

\textit{Example.}  
For $\delta=0.01$ and $\epsilon=0.05$, Hoeffding’s bound requires 
$N \approx 85.5\times 10^{3}$ copies to achieve confidence $0.99$.  
If a higher error probability $\delta=0.05$ is acceptable, Chebyshev’s bound suffices with only $N \approx 24\times 10^{3}$.  
Suppose the measured value is $\Bar{S}=2\sqrt{2}-0.03$; then the corresponding high-confidence interval for the entanglement fidelity is $[0.972,\,1]$.
% \subsection{\textbf{Statistical Verification}}

% Building on our statistical estimation framework, we now introduce a hypothesis testing approach for entanglement verification. This method allows us to make binary decisions about the quality of entanglement in quantum networks with quantifiable statistical confidence. For a quantum state $\rho\in \mathcal{Q}$ and small parameter $\alpha \in (0,1)$, we design a test to distinguish two hypotheses:
% \begin{equation}
%    H_0: F(\rho) \geq 1-\alpha \quad \text{versus}\quad H_1: F(\rho) \leq 1 -3\alpha.
%    \label{hypothesis with a gap}
% \end{equation}

\subsection{\textbf{Statistical Verification}}

Building on the statistical estimation framework, we introduce a hypothesis-testing approach for entanglement verification. This method enables binary decisions on entanglement quality in quantum networks with quantifiable confidence. For a state $\rho\in\mathcal{Q}$ and parameter $\alpha\in(0,1)$, we test:
\begin{equation}
   H_0: F(\rho) \geq 1-\alpha 
   \quad \text{vs.} \quad 
   H_1: F(\rho) \leq 1-3\alpha,
   \label{hypothesis with a gap}
\end{equation}
where the gap between $H_0$ and $H_1$ provides robustness against statistical fluctuations.

\begin{algorithm}
\caption{CHSH-based Entanglement Verification (EV)}
\begin{algorithmic}[1]
\STATE \textbf{Input:}  $\delta \in (0,1)$, and $\alpha \in (0,1/3)$
\STATE Define $N = \min \left\{\frac{3}{\delta\alpha^2}, \frac{16}{\alpha^2}\ln{\frac{2}{1-(1-\frac{\delta}{2})^{\frac{1}{4}}}}\right\}$ \\and $\theta_{th} = 2\sqrt{2}-5\sqrt{2}\alpha$
\STATE Run Algorithm \ref{Alg: measure CHSH} with input $N$ and obtain $\bar{S}$
\IF{$\bar{S}\geq\theta_{th}$}
   \STATE \textbf{Return:} Accept as $H_0$
\ELSIF{$\bar{S} < \theta_{th}$}
   \STATE \textbf{Return:} Reject as $H_1$
\ENDIF
\end{algorithmic}
\label{Alg: Verification}
\end{algorithm}

% Algorithm \ref{Alg: Verification}, which is called \textit{CHSH-based Entanglement Verification} (EV), provides a practical implementation of this test based on our CHSH measurement protocol. By selecting an appropriate threshold $\theta_{th}$ and input $N$ for Algorithm \ref{Alg: measure CHSH}, i.e. sample size $4N$, we can distinguish between these hypotheses with error probability bounded by $\delta$.

% \begin{theorem}
% Protocol CHSH-based Entanglement Verification (EV) successfully distinguishes $H_0$ and $H_1$ with probability at least $1-\delta$.
%     \label{theorem: performance of verification protocol}
% \end{theorem}

Algorithm~\ref{Alg: Verification}, named as \textit{CHSH-based Entanglement Verification} (EV), implements the hypothesis test in \eqref{hypothesis with a gap}. By choosing the threshold $\theta_{th}$ and sample size $4N$ as specified, the test accepts or rejects with error probability at most $\delta$.

\begin{theorem}
The CHSH-based Entanglement Verification (EV) protocol distinguishes between $H_0$ and $H_1$ with probability at least $1-\delta$.
\label{theorem: performance of verification protocol}
\end{theorem}

% \textit{Proof.} We bound the error probability of Algorithm \ref{Alg: Verification} while denoting $S=S(\rho)$ and $F=F(\rho)$ for short (see Appendix A for the detailed derivation)
% \begin{align}
%     \mathbb{P}[\text{Error}]&= \mathbb{P}[\bar{S}\geq\theta_{th},F\leq 1-3\alpha]+\mathbb{P}[\bar{S}<\theta_{th},F\geq 1-{\alpha}]\notag\\
%      &\leq 2\mathbb{P}[|\bar{S}-S|\geq\sqrt{2}\alpha].
% \end{align}
% This error probability can be bounded using concentration inequalities. Applying Hoeffding's inequality yields:
% \begin{align}
%     \mathbb{P}[\text{Error}]\leq 2\left[1-\left(1-2\exp{\left(-\frac{N\alpha^2}{16}\right)}\right)^4\right].
% \end{align}
% Alternatively, using Chebyshev's inequality gives:
% \begin{align}
%     \mathbb{P}[\text{Error}]\leq \frac{3}{N\alpha^2}.
% \end{align}

% Therefore, by selecting the input for Algorithm \ref{Alg: measure CHSH} as $N = \min \left\{\frac{3}{\delta\alpha^2}, \frac{16}{\alpha^2}\ln{\frac{2}{1-(1-\frac{\delta}{2})^{\frac{1}{4}}}}\right\}$, we ensure that the error probability is bounded above by $\delta$, completing the proof.\qed

\textit{Proof.} Denote $S=S(\rho)$ and $F=F(\rho)$ for short.  
The error probability of Algorithm~\ref{Alg: Verification} is (see Supplementary Material A for details)
\begin{align}
    \mathbb{P}[\text{Error}]
        &= \mathbb{P}[\Bar{S}\geq \theta_{th},\,F\leq 1-3\alpha]
         + \mathbb{P}[\Bar{S}< \theta_{th},\,F\geq 1-\alpha] \notag\\
        &\leq 2\,\mathbb{P}[\,|\Bar{S}-S|\geq \sqrt{2}\alpha\,].
\end{align}
This probability can be bounded via concentration inequalities.  
Applying Hoeffding’s inequality gives
\begin{align}
    \mathbb{P}[\text{Error}]
        &\leq 2\Bigl[1-\bigl(1-2\exp(-N\alpha^2/16)\bigr)^4\Bigr],
\end{align}
while Chebyshev’s inequality yields
\begin{equation}
    \mathbb{P}[\text{Error}] \leq \frac{3}{N\alpha^2}.
\end{equation}
Therefore, by selecting the input for Algorithm \ref{Alg: measure CHSH} as $N = \min \left\{\frac{3}{\delta\alpha^2}, \frac{16}{\alpha^2}\ln{\frac{2}{1-(1-\frac{\delta}{2})^{\frac{1}{4}}}}\right\}$, we ensure that the error probability is bounded above by $\delta$, completing the proof.\qed

% This verification approach is particularly valuable in quantum network settings where binary decisions about entanglement quality must be made efficiently. Theoretically, our approach sets an asymptotic upper bound for the required computing resources at $O(poly(\alpha,\log (\frac{1}{\delta})))$ while maintaining quantifiable statistical guarantees.

This verification approach is particularly valuable for quantum networks where binary decisions on entanglement quality must be made efficiently. It yields an asymptotic upper bound on the required resources of $O(\mathrm{poly}(1/\alpha,\,\log(1/\delta)))$ while maintaining rigorous statistical guarantees.

% \subsection{\textbf{Robustness and Limitation}}

% The robustness of using CHSH violation to lower-bound the entanglement fidelity arises from the fact that the gap between the true fidelity $F(\rho)$ and the lower bound $S(\rho)/(2\sqrt{2})$ depends solely on the $|\Psi^{-}\rangle$ population in the state. In many physical settings—particularly those relevant for photonic entanglement sources—this population corresponds to a specific subset of Pauli errors, namely effective $Y$ errors (including correlated $X$ and $Z$ errors on different qubits). Such events can often be engineered to be rare: for example, in the photonic ``cluster state machine gun'' proposal of Lindner and Rudolph~\cite{Lindner2009}, $Y$-type errors originate primarily from spin precession during photon emission, which can be strongly suppressed through low magnetic fields, phase-correction pulses, and spectral filtering. Their analysis indicates that with realistic parameters, the per-photon Pauli error probabilities can be reduced below $0.2\%$, with $Y$ errors forming only a small fraction of the total. Under these conditions, the lower bound remains tight in practical regimes, with the true fidelity only marginally higher. 

\subsection{\textbf{Robustness and Limitation}}

The robustness of using CHSH violation to lower-bound entanglement fidelity stems from the fact that the gap between the true fidelity $F(\rho)$ and its lower bound $S(\rho)/(2\sqrt{2})$ depends solely on the $|\Psi^{-}\rangle$ population in the state. In many physical settings—particularly those involving photonic entanglement sources—this population corresponds to a specific subset of Pauli errors, namely effective $Y$ errors (including correlated $X$ and $Z$ errors on different qubits). Such events can often be engineered to be rare. For example, in the photonic “cluster state machine gun’’ proposal of Lindner and Rudolph~\cite{Lindner2009}, $Y$-type errors arise mainly from spin precession during photon emission, which can be strongly suppressed using low magnetic fields, phase-correction pulses, and spectral filtering. Their analysis shows that under realistic parameters, per-photon Pauli error probabilities can be reduced below $0.2\%$, with $Y$ errors forming only a small fraction of the total. In this regime, the CHSH-based lower bound remains practically tight, with the true fidelity only marginally higher.

The limitation, however, is fundamental: CHSH violation provides only a sufficient condition for entanglement. When $S(\rho)$ is well below the quantum maximum $2\sqrt{2}$, the gap between lower and upper bounds widens regardless of experimental optimization. In low-quality entanglement regimes, or when $Y$-type errors are not negligible, the lower bound may substantially underestimate fidelity, reducing the protocol’s decisiveness.

% \begin{figure*}[!t]
%     \centering
%     % --- Row 1 ---
%     \subfloat[Depolarizing (Werner)\label{fig:depol}]{
%         \includegraphics[width=0.4\textwidth]{figs/werner_seed100.png}
%     }\hfill
%     \subfloat[Amplitude damping\label{fig:ampdamp}]{
%         \includegraphics[width=0.4\textwidth]{figs/amplitude_damping_seed100.png}
%     }

%     \vspace{0.6em}

%     % --- Row 2 ---
%     \subfloat[Phase damping\label{fig:phasedamp}]{
%         \includegraphics[width=0.4\textwidth]{figs/phase_damping_seed100.png}
%     }\hfill
%     \subfloat[Dephasing (Pauli Z)\label{fig:dephasing}]{
%         \includegraphics[width=0.4\textwidth]{figs/dephasing_seed100.png}
%     }

%     \caption{Performance comparison between CHSH-based and $n{=}2$ GHZ-state verification protocols under four noise models. }
%     \label{fig:noise_comparison}
% \end{figure*}

% \begin{figure*}[!t]
%     \centering
%     \subfloat[Depolarizing\label{fig:depol}]{
%         \includegraphics[width=0.3\textwidth]{Figures/GHZ_comparison/werner_seed100.png}
%     }\hfill
%     \subfloat[Amplitude damping\label{fig:ampdamp}]{
%         \includegraphics[width=0.3\textwidth]{Figures/GHZ_comparison/amplitude_damping_seed100.png}
%     }\hfill
%     \subfloat[Dephasing\label{fig:dephasing}]{
%         \includegraphics[width=0.3\textwidth]{Figures/GHZ_comparison/dephasing_seed100.png}
%     }
%     \caption{Comparison of CHSH-based and $n{=}2$ GHZ-state verification protocols under three noise models: depolarizing, amplitude damping, and dephasing.}
%     \label{fig:noise_comparison}
% \end{figure*}

\begin{figure*}[!t]
    \centering
    \subfloat[\label{fig:depol}]{
        \includegraphics[width=0.3\textwidth]{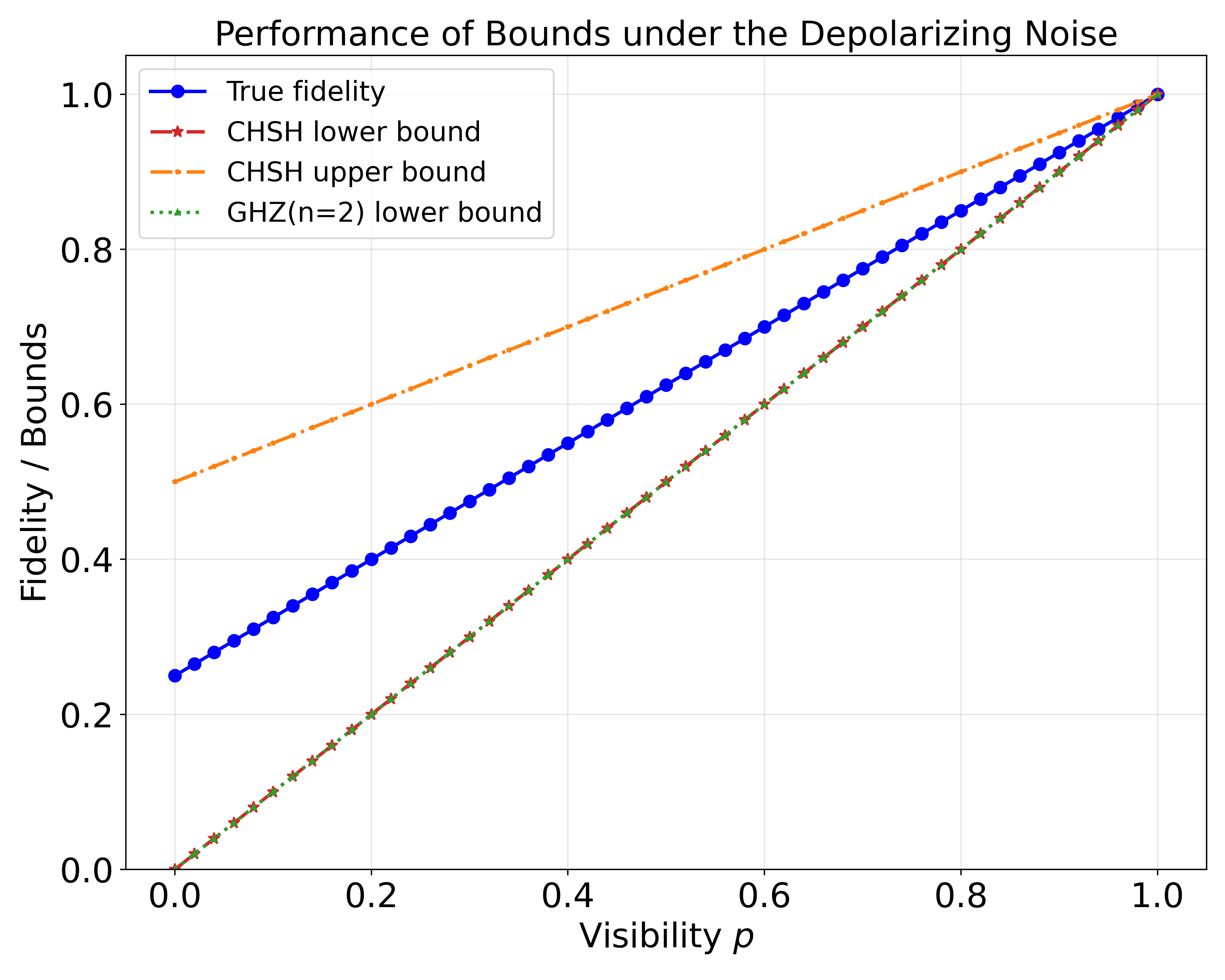}
    }\hfill
    \subfloat[\label{fig:ampdamp}]{
        \includegraphics[width=0.3\textwidth]{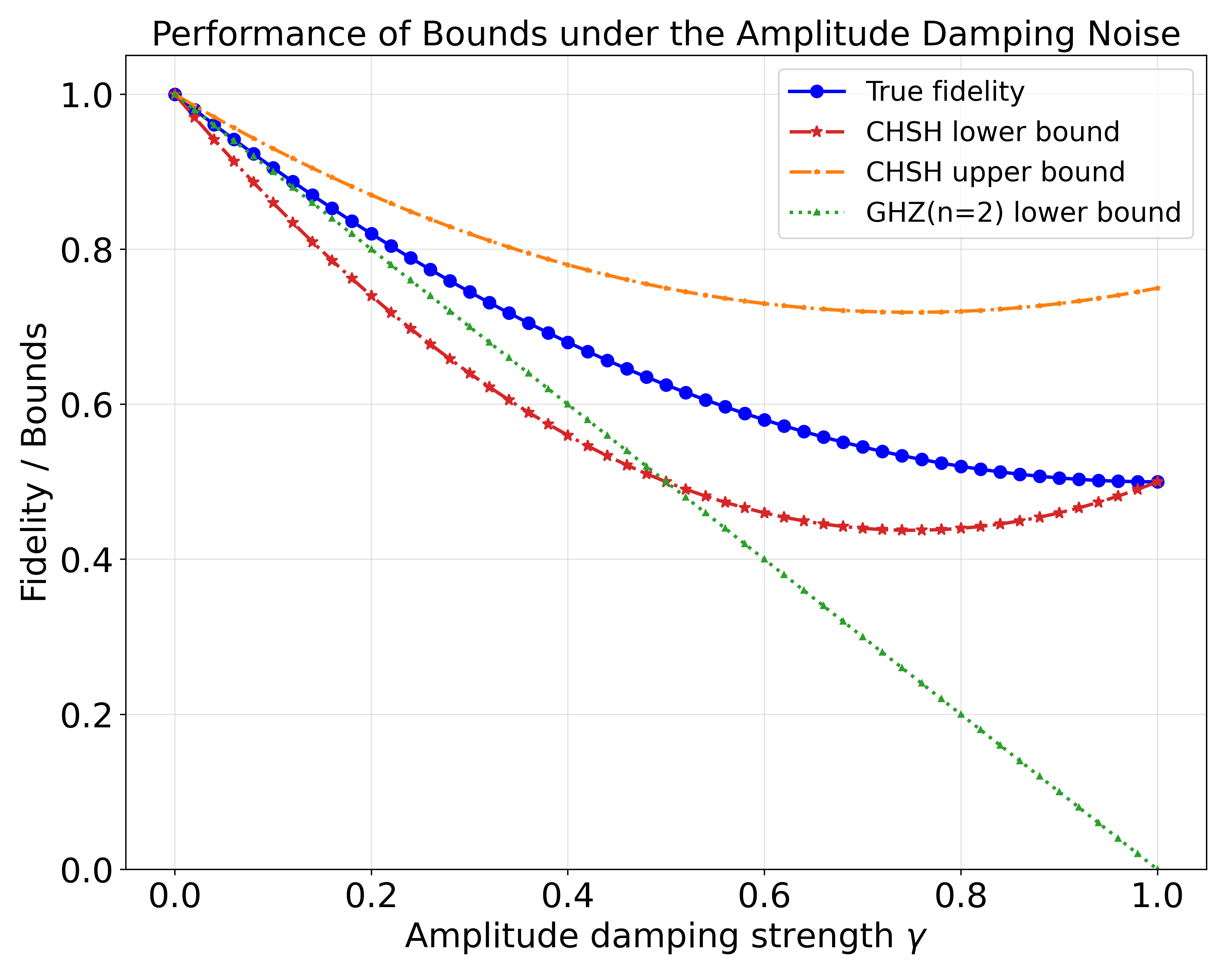}
    }\hfill
    \subfloat[\label{fig:dephasing}]{
        \includegraphics[width=0.3\textwidth]{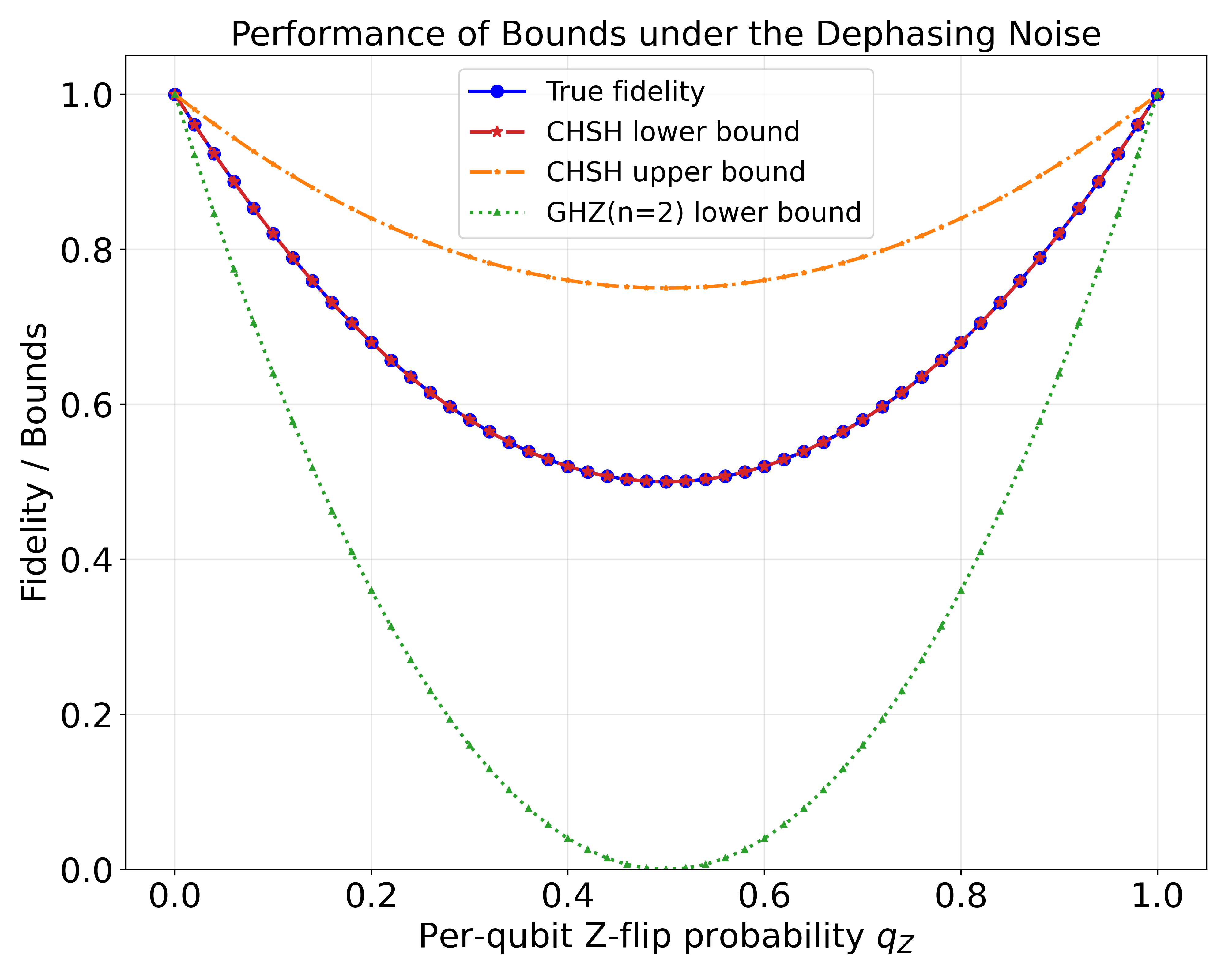}
    }
    \caption{Comparison of CHSH-based and $n{=}2$ GHZ-state verification protocols under noise models: (a) depolarizing, (b) amplitude damping, and (c) dephasing.}
    \label{fig:noise_comparison}
\end{figure*}

\subsection{\textbf{Comparison to Existing Methods}}

\textit{Direct Fidelity Estimation.}  
For the EPR target $|\Phi^+\rangle$, DFE reduces to measuring three signed Pauli correlators,
\[
\hat{F}_{\mathrm{DFE}} = \tfrac14\!\left(1 + \langle XX\rangle_\rho - \langle YY\rangle_\rho + \langle ZZ\rangle_\rho\right).
\]
% Since $|\Phi^+\rangle$ is a stabilizer state, these settings occur with equal probability, making the estimation complexity independent of Hilbert space dimension~\cite{flammia2011direct}. 

By Hoeffding’s inequality, the copy complexity for testing $H_0: F \ge 1-\alpha$ versus $H_1: F \le 1-3\alpha$ with symmetric risk $\delta$ and threshold $t = 1-2\alpha$ is $\mathbb{E}[M] = \tfrac{8}{\alpha^2}\ln\!\tfrac{2}{\delta}$ (see Supplementary Material B). Our CHSH-based test achieves the same asymptotic scaling for additive-error estimation and symmetric-gap fidelity testing, with four correlators instead of three Pauli settings. The CHSH approach additionally links nonlocality to fidelity: any DFE fidelity bound can be translated into a CHSH statement; in particular, $F > \tfrac{1+\sqrt{2}}{2\sqrt{2}} \approx 0.8536$ implies $S > 2$ and hence nonlocality. The key advantage of the single statistic $S$ is that it directly certifies Bell inequality violation, enabling device-independent (DI) protocols, whereas DFE assumes trusted, calibrated measurements and cannot directly provide DI guarantees. While our analysis focuses on $|\Phi^+\rangle$ for simplicity, the method extends to any maximally entangled state via local unitaries, and the diagonalization of $\hat{S}$ might be applied to other Bell-type inequalities to derive both upper and lower fidelity bounds.

\textit{Existing GHZ Verification Protocol.}  
For $n=2$, the GHZ-state verification protocol~\cite{McCutcheon2016-GHZ-verification}, which in this case reduces to an EPR test, estimates a fidelity lower bound from equatorial-basis correlations. Under simple depolarizing noise (Fig.~\ref{fig:depol}), this produces the same result as the analytical CHSH lower bound. Important differences arise, however. First, our CHSH-based protocol also provides an \emph{upper bound} on $F(\rho)$, preventing overestimation of fidelity, a feature absent in the GHZ method. Second, while both approaches behave similarly under depolarizing noise, simulations show that the CHSH bound remains much tighter under strong amplitude damping (Fig.~\ref{fig:ampdamp}), where the GHZ bound quickly becomes loose. This occurs because amplitude damping populates states outside the equatorial subspace probed by GHZ measurements, whereas CHSH settings still capture the relevant correlations. The largest gap appears under pure dephasing (Fig.~\ref{fig:dephasing}), where the GHZ method performs poorly: dephasing destroys the phase coherence that the GHZ test relies on, whereas CHSH violation remains robust since such channels do not introduce effective $Y$ errors. The definitions of these noise models and related analysis are given in Supplementary Material C.

\subsection{\textbf{Practical Entanglement Verification}}

Consider a practical scenario where Alice and Bob wish to teleport quantum information, which requires high-quality EPR states. They adopt a quantitative security criterion: teleportation proceeds only if the entanglement fidelity exceeds $1-\alpha$, with $\alpha>0$ small. This motivates the need for a reliable verification process.

Using CHSH measurements, one can estimate entanglement fidelity with statistical confidence, but an inherent limitation arises. As shown in Theorem~\ref{theorem: bound entanglement fidelity} and Corollary~\ref{Cor: Estimate Fidelity using chsh}, CHSH provides both upper and lower bounds on fidelity, yet a non-negligible gap always remains. This gap grows as the CHSH value $S(\rho)$ moves away from the quantum maximum $2\sqrt{2}$. Theoretically, this gap partially explains why Algorithm~\ref{Alg: Verification} distinguishes only between hypotheses $H_0$ and $H_1$ separated by $2\alpha$, leaving an ambiguous region $(1-3\alpha,\,1-\alpha)$ where no definitive conclusion can be drawn.

While the gap can be reduced by choosing smaller $\alpha$, this confines the protocol to extremely high-fidelity regimes, effectively accepting only near-perfect entanglement, which is impractical in real-world settings. To address this, we propose a modified protocol in Algorithm~\ref{Alg: Practical Verification}, termed \textit{Practical CHSH-based Entanglement Verification} (PEV), which directly tests
\begin{equation*}
   H_0: F(\rho) \geq 1-\alpha 
   \quad \text{versus} \quad 
   H_1: F(\rho) < 1-\alpha.
\end{equation*}

The protocol consumes $4N$ EPR pairs, allocating $N$ pairs per correlation term $\langle A_iB_j \rangle$ via Algorithm~\ref{Alg: measure CHSH}, to produce an estimate $\Bar{S}$ of $S(\rho)$. It accepts $H_0$ if $\Bar{S}\geq \theta_{th}$ and rejects as $H_1$ otherwise. The threshold
$\theta_{th} = 2\sqrt{2} - \tfrac{5\sqrt{2}}{3}\alpha$ is inspired from the EV protocol and the fidelity bound in Theorem~\ref{theorem: bound entanglement fidelity}, chosen to balance success probability with sample complexity.

% More specifically, we decrease the lower bound of $H_0$ from $ \left[ F (\rho) \geq 1-\alpha \right] \rightarrow \left [ F (\rho) \geq 1-3\alpha\right]$ to close the gap of width $2\alpha$, then rescale $3\alpha \rightarrow \alpha$.

\begin{algorithm}
\caption{Practical CHSH-based Entanglement Verification (PEV)}
\begin{algorithmic}[1]
\STATE \textbf{Input:}  $\delta \in (0,1)$, and $\alpha \in (0,1/3)$
\STATE Define $N = \min \left\{\frac{3}{\delta\alpha^2}, \frac{16}{\alpha^2}\ln{\frac{2}{1-(1-\frac{\delta}{2})^{\frac{1}{4}}}}\right\}$ \\and $\theta_{th} = 2\sqrt{2}-\frac{5\sqrt{2}}{3}\alpha$
\STATE Run Algorithm \ref{Alg: measure CHSH} with input $N$ and obtain $\bar{S}$
\IF{$\bar{S}\geq\theta_{th}$}
   \STATE \textbf{Return:} Accept as $H_0$
\ELSIF{$\bar{S} < \theta_{th}$}
   \STATE \textbf{Return:} Reject as $H_1$
\ENDIF
\end{algorithmic}
\label{Alg: Practical Verification}
\end{algorithm}

% \begin{algorithm}
% \caption{Practical CHSH-based Entanglement Verification (PEV)}
% \begin{algorithmic}[1]
% \STATE \textbf{Input:}  $N$, and $\alpha \in (0,1)$
% \STATE Define  $\theta_{th} = 2\sqrt{2}-\frac{5\sqrt{2}}{3}\alpha$
% \STATE Run Algorithm \ref{Alg: measure CHSH} with input $N$ and obtain $\bar{S}$
% \IF{$\bar{S}\geq\theta_{th}$}
%    \STATE \textbf{Return:} Accept as $H_0$
% \ELSIF{$\bar{S} < \theta_{th}$}
%    \STATE \textbf{Return:} Reject as $H_1$
% \ENDIF
% \end{algorithmic}
% \label{Alg: Practical Verification}
% \end{algorithm}
In the following section, we evaluate our method with protocol PEV through comprehensive simulations of quantum networks. These simulations demonstrate the practical performance of our theoretical framework and illustrate its effectiveness under various network conditions.
\section{Evaluation Methodology and Implementation}

\subsection{\textbf{NetSquid Framework}}

% NetSquid (Network Simulator for Quantum Information using Discrete Events) is a discrete-event simulation framework designed specifically for simulation of quantum networks and communication protocols under realistic conditions \cite{Coopmans2021-NetSquid}. It provides a high-performance platform capable of modeling modular quantum computing systems across multiple abstraction layers—from detailed physical hardware components (such as photon sources, quantum memories, and detector assemblies) to complex application-level protocols (including quantum money, verifiable blind quantum computation, and quantum digital signatures) \cite{liao2022benchmarking}.  

NetSquid (Network Simulator for Quantum Information using Discrete Events) is a discrete-event framework tailored for simulating quantum networks and communication protocols under realistic conditions \cite{Coopmans2021-NetSquid}. It provides a platform for modeling modular quantum systems across multiple abstraction layers—from physical hardwares (e.g., photon sources, quantum memories, detectors) to application-level protocols (e.g., quantum money, verifiable blind computation, quantum digital signatures) \cite{liao2022benchmarking}.

% A key strength of NetSquid's architecture lies in its ability to accurately capture time-dependent quantum phenomena (including qubit decoherence processes and entanglement lifetime limitations) alongside critical network-level effects (such as communication delays, transmission losses, and precise control timing). This enables accurate simulation of both noisy quantum operations and the classical signaling mechanisms necessary for protocol execution. Given these capabilities, we employ NetSquid as the foundation for constructing our quantum network simulator, which allows us to rigorously evaluate and systematically optimize the performance characteristics of protocol PEV.

A key strength of NetSquid is its ability to capture time-dependent quantum effects (e.g., decoherence and entanglement lifetime) together with network-level phenomena (e.g., delays, transmission losses, and control timing). This allows realistic simulation of both noisy quantum operations and the classical signaling required for protocol execution. Leveraging these features, we use NetSquid as the foundation of our simulator to evaluate and optimize the performance of protocol PEV.

\textit{Framework Contribution.}  
The contribution of our framework lies not only in fixing a single verification protocol, but in providing a general methodology for simulating and evaluating CHSH-based procedures. Because the physical noise affecting photons in long-distance channels is not yet fully understood, perfectly realistic simulations are not possible. Instead, our modular design allows researchers to configure noise models, hardware assumptions, and resource constraints as needed. Thus, the simulator serves as a versatile tool for designing and testing a broad class of CHSH-based protocols, offering insights adaptable to diverse experimental and network settings.

\subsection{\textbf{Implementation \& Setup}}
% Our simulation methodology begins with a quantum network architecture, followed by specialized modules that implement CHSH-based entanglement verification protocols.

% \noindent
% \textbf{Network Architecture.} The quantum network model comprises the following essential components:
% \begin{itemize}
%     \item \textit{Network Topology:} A two-node configuration (Alice and Bob) connected via a noisy quantum channel for entanglement distribution and a classical channel for exchanging measurement basis information, outcomes, and verification results.
    
%     \item \textit{Quantum Memory and Processors:} Both nodes incorporate quantum memory with configurable decoherence properties, enabling precise modeling of quantum state degradation over time. The memory stores multiple qubits and supports the quantum operations required for measurement and teleportation.
    
%     \item \textit{Entanglement Generation:} A configurable entanglement source designed to distribute EPR pairs to Alice and Bob.
    
%     \item \textit{Verification Protocol:} An implementation of protocol PEV to verify the quality of distributed entanglement.
    
%     \item \textit{Quantum Teleportation Protocol:} A module that executes quantum teleportation operations contingent upon successful verification.
% \end{itemize}

Our simulation methodology begins with a quantum network architecture, followed by specialized modules that implement CHSH-based entanglement verification protocols.

\noindent
\textbf{Network Architecture.} The model includes:

\begin{itemize}
    \item \textit{Network Topology:} Two nodes (Alice and Bob) connected by a noisy quantum channel for entanglement distribution and a classical channel for exchanging bases, outcomes, and verification results.
    
    \item \textit{Quantum Memory and Processors:} Configurable memories with decoherence properties, capable of storing multiple qubits and supporting operations for measurement and teleportation.
    
    \item \textit{Entanglement Generation:} A source distributing EPR pairs to Alice and Bob.
    
    \item \textit{Verification Protocol:} Entanglement quality verification.
    
    \item \textit{Quantum Teleportation Protocol:} Executes teleportation once verification succeeds.
\end{itemize}
While our simulator supports multi-node architectures and entanglement swapping, in this paper we restrict attention to the simplest two-node configuration in order to focus on entanglement verification.
\begin{figure}[!t]
    \centering
    \includegraphics[width=0.8\linewidth]{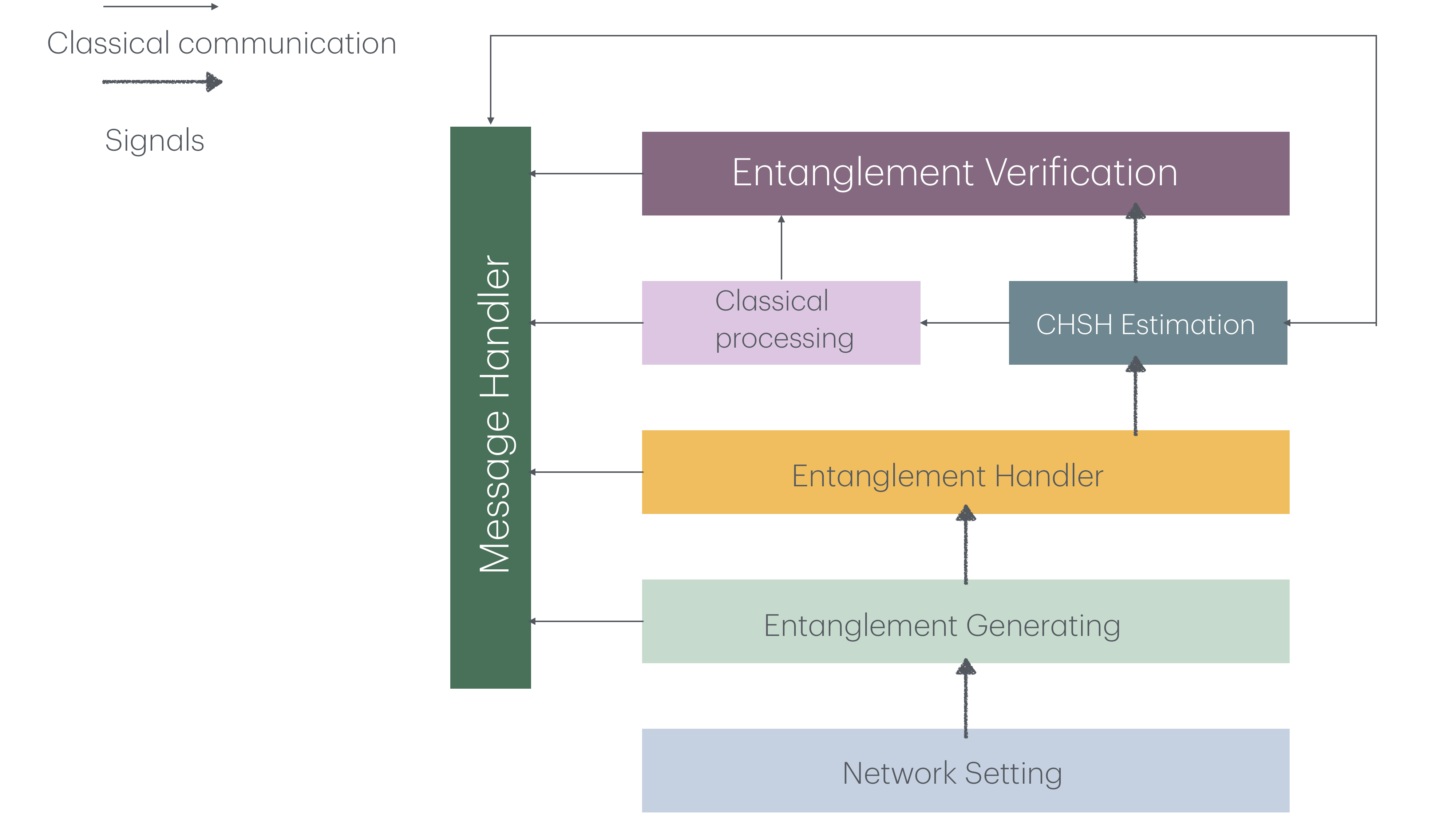}
    \caption{Schematic of the network model implementing CHSH-based verification. Other CHSH-based protocols can also be integrated.}
    \label{fig:CHSH_implementation}
\end{figure}

\noindent
\textbf{Implementation Structure.} As illustrated in Figure~\ref{fig:CHSH_implementation}, our CHSH-based verification protocol is implemented through seven interconnected modular components:

\begin{itemize}
\item \textit{Entanglement Generating Module:} Produces EPR pairs, with evolution and noise determined by network settings.

\item \textit{Entanglement Handler Module:} Tracks generated pairs, memory use, updating as new entanglement is created.

\item \textit{Classical Processing Module:} Configures CHSH protocols by coordinating basis, selecting pairs, and collecting results.

\item \textit{CHSH Estimation Module:} Executes Algorithm~\ref{Alg: measure CHSH} as directed by the Classical Processing Module.

\item \textit{Entanglement Verification Module:} Compares $\Bar{S}$ with threshold $\theta_{th}$ to decide between $H_0$ and $H_1$.

\item \textit{Message Handler Module:} Handles classical messages on basis exchange, outcome reporting, and verification.
\end{itemize}

\subsection{\textbf{Evaluation Framework}}

% \textbf{Network Configuration (Physical Model).}
% In our NetSquid implementation, the physical model of the fiber-based quantum network can be adjusted directly. The classical and quantum channels are modeled with parameters for transmission delay, fiber loss per kilometer, and depolarizing noise. Entangled pairs are generated by a source with a configurable emission rate and adjustable state distribution. Each node also contains a quantum memory whose size and noise level can be set, allowing us to capture how qubits degrade over time.

% Concretely, our code parameterizes: 
% (i) the physical separation $L$ (km), which together with the propagation velocity $v$ in the FibreDelayModel determines the channel delay; 
% (ii) the fiber attenuation coefficient $\mu$; 
% (iii) the quantum channel depolarization rate $R_c$ (Hz); 
% (iv) the quantum memory size and memory depolarization rate $R_m$; 
% and (v) the source emission period $\tau_{\mathrm{src}}$. These parameters can be modified directly in the simulation, allowing us to sweep over distance, fiber quality, channel noise, memory quality, and source brightness, or to substitute alternative noise models. This modular structure makes the simulator adaptable to a broad range of experimental assumptions.

\textbf{Physical configuration.}
In our NetSquid implementation, the fiber-based network is parameterized directly. Classical and quantum channels include transmission delay, fiber loss per kilometer, and noise model. Entangled pairs are produced by a source with configurable emission rate and state distribution. Each node hosts a quantum memory with adjustable size and decoherence rate.

Specifically, we model: 
(i) separation $L$ and propagation velocity $v$ (channel delay); 
(ii) fiber attenuation $\mu$; 
(iii) channel depolarization rate $R_c$; 
(iv) memory size and depolarization rate $R_m$; 
(v) source emission period $\tau_{\mathrm{src}}$. 
These parameters can be tuned to explore distance, fiber quality, noise, memory fidelity, and source brightness, or replaced with alternative noise models. This modular design ensures adaptability across experimental settings.

\begin{table}[h]
\centering
\caption{Baseline Configuration Parameters}
\label{tab:baseline}
\begin{tabular}{|l|c|c|}
\hline
\textbf{Parameter} & \textbf{Symbol} & \textbf{Value/Range} \\
\hline
Fidelity threshold parameter & $\alpha$ & 0.1 \\
Total EPR pair capacity & $\mathcal{C}$ & 10{,}000 \\
Verification sample fraction & $\beta$ & 0.3 \\
Acceptable failure rate & $\delta$ & 0.1 \\
Node separation distance & $L$ & 1 km \\
Propagation speed in fiber & $v$ & $2.0\times 10^{8}$ m/s \\
Channel depolarization rate & $R_c$ & 8,000 Hz \\
Memory depolarization rate & $R_m$ & 10 Hz \\
Fiber attenuation & $\mu$ & 0.2 dB/km \\
Source emission period & $\tau_{\mathrm{src}}$ & $1\times 10^{5}$ ticks \\
\hline
\end{tabular}
\end{table}

% \textbf{Baseline Configuration.} 
% We evaluate protocol PEV across various operational conditions to assess parameter sensitivity and identify efficiency regimes. The baseline configuration is defined by the parameters in Table~\ref{tab:baseline}. In summary, teleportation is attempted only when the estimated entanglement fidelity exceeds $1-\alpha$, subject to a resource limit of $\mathcal{C}$ EPR pairs and an acceptable failure rate $\delta$. From the total pool, a fraction $\beta\mathcal{C}$ is allocated to CHSH estimation, while the remaining $(1-\beta)\mathcal{C}$ pairs are reserved for teleportation if verification succeeds. 

% Because PEV is inherently probabilistic, increasing the sample size improves verification accuracy. We therefore define the optimal sampling fraction $\beta^*$ as the minimum value that ensures the protocol’s success rate meets or exceeds $1-\delta$. This balances the trade-off between verification confidence and the number of pairs left for teleportation.
% By default, all simulations use the fixed parameters in the baseline. For specific experiments, one parameter is varied, with the corresponding ranges provided in the experiment descriptions. For statistical robustness, each data point reported in the figures represents the average outcome of 200 independent repetitions of protocol PEV under identical conditions.

\textbf{Baseline Configuration.} 
Protocol PEV is evaluated under varying conditions to assess parameter sensitivity and efficiency. The baseline setup (Table~\ref{tab:baseline}) attempts teleportation only when the estimated fidelity exceeds $1-\alpha$, with a limited resource of $\mathcal{C}$ EPR pairs and tolerated failure $\delta$. A fraction $\beta\mathcal{C}$ is used for CHSH estimation, leaving $(1-\beta)\mathcal{C}$ pairs for teleportation if verification succeeds.  

Since PEV is inherently probabilistic, larger samples improve accuracy. We define the optimal sampling fraction $\beta^*$ as the smallest value ensuring success probability at least $1-\delta$, balancing verification confidence with teleportation resources. All simulations use the baseline parameters unless noted, with one parameter varied at a time. Each data point reflects the mean of 200 independent runs for statistical robustness.

% \textbf{Performance Metrics.} Using NetSquid's capability to track quantum states throughout the simulation, we can compute the mean value of actual entanglement fidelity $F$ for the remaining $(1-\beta)\mathcal{C}$ EPR pairs in quantum memory. This allows us to determine whether hypothesis $H_0: F \geq 0.9$ or $H_1: F < 0.9$ truly holds for these remaining pairs, providing ground truth for evaluating whether PEV's decision (based only on the $\beta\mathcal{C}$ sampled pairs) was correct.
% % \textit{Verification Accuracy}: 
% We quantify accuracy through false positive rate (FPR) and false negative rate (FNR):
% \begin{align}
% \text{FPR} &= \frac{\text{\# of cases where PEV accepts when } H_1 \text{ holds}}{\text{Total number of repetitions}} \\
% \text{FNR} &= \frac{\text{\# of cases where PEV rejects when } H_0 \text{ holds}}{\text{Total number of repetitions}}
% \end{align}

% The overall success rate of the protocol is then defined as:
% \begin{align}
% \text{Success Rate} = \frac{\text{\# of cases where PEV decides correctly}}{\text{Total number of repetitions}}.
% \end{align}

\textbf{Performance Metrics.} 
NetSquid tracks quantum states, allowing us to compute the mean fidelity $F$ of the remaining $(1-\beta)\mathcal{C}$ EPR pairs. This gives ground truth for testing whether $H_0: F \geq 0.9$ or $H_1: F < 0.9$, and for checking if PEV’s decision (based only on $\beta\mathcal{C}$ sampled pairs) is correct.

We measure accuracy via false positive (FPR) and false negative (FNR) rates:
\begin{align}
\text{FPR} &= \frac{\#\{\text{PEV accepts while } H_1 \text{ holds}\}}{\text{Total repetitions}}, \\
\text{FNR} &= \frac{\#\{\text{PEV rejects while } H_0 \text{ holds}\}}{\text{Total repetitions}}.
\end{align}

The success rate is:
\begin{align}
\text{Success Rate} = \frac{\#\{\text{PEV decides correctly}\}}{\text{Total repetitions}}.
\end{align}

% (2) \textit{Resource Efficiency}: This is determined by the minimum required verification sample fraction $\beta^*$ that achieves our target success rate:
% \begin{align}
% \beta^* = \min\{\beta \in [0.1, 0.7] \mid \text{Success Rate}(\beta) \geq 1-\delta\}
% \end{align}

% These metrics enable us to identify the optimal trade-off between verification confidence and maximizing the number of EPR pairs available for teleportation.

% \noindent
\textbf{Experimental Variations.} To comprehensively assess protocol robustness, we conduct four distinct sets of experiments by varying individual parameters while holding others constant, as detailed in Table~\ref{tab:experiments}.
\begin{table}[h]
\centering
\caption{Experimental Parameter Variations}
\label{tab:experiments}
\begin{tabular}{|p{2.5cm}|p{3cm}|p{1.8cm}|}
\hline
\textbf{Experiment} & \textbf{Variable Parameter} & \textbf{Range} \\
\hline
Sample Size Effect & Sample fraction ($\beta$) & 0.1 to 0.7 \\
\hline
Distance Sensitivity & Node separation ($L$) & 0.5 to 3 km \\
\hline
Channel Quality & Quantum channel depolarization rate ($R_c$) & 1,000 to 16,000 Hz \\
\hline
Threshold Sensitivity & Fidelity threshold ($\alpha$) & 0.01 to 0.20 \\
\hline
\end{tabular}
\end{table}

\subsection{\textbf{Results}}
% \textbf{PEV Performance with Varying Sample Size.} 
% This experiment examines how verification accuracy changes with different sample fractions ($\beta$). We maintain fixed values for depolarization rates ($R_c = 8000$ Hz), node distance ($L = 1$ km), and threshold ($\alpha = 0.1$) while increasing the fraction $\beta$ of EPR pairs allocated to verification by $0.1$ at each step. Figure \ref{fig:sample-size-performance} shows both the success rate and error composition as functions of $\beta$. 

\begin{figure} [h]
    \centering
    \includegraphics[width=0.6\linewidth]{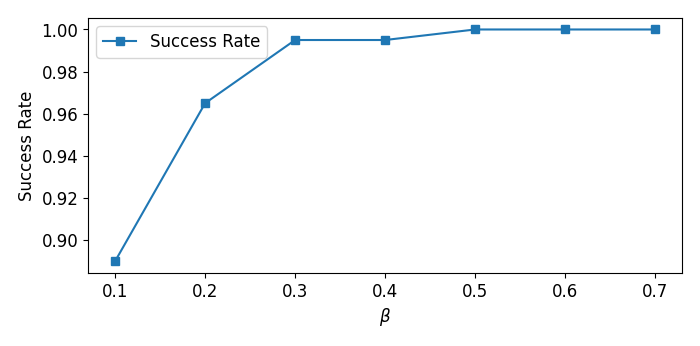}
    \includegraphics[width=0.6\linewidth]{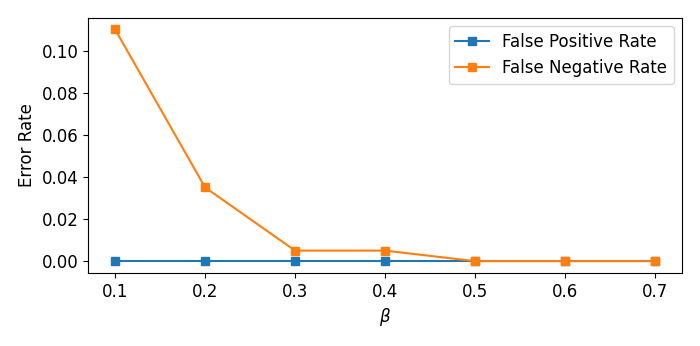}
    \caption{PEV performance as a function of sampling fraction $\beta$. Top: Overall success rate with $\mathcal{C}=10k$ and $\alpha =0.1$.  Bottom: Decomposition of errors into false positives and false negatives.}
    \label{fig:sample-size-performance}
\end{figure}

\textbf{PEV Performance with Varying Sample Size.} 
We study how verification accuracy depends on the sample fraction $\beta$. Depolarization rate ($R_c=8000$ Hz), distance ($L=1$ km), and threshold ($\alpha=0.1$) are fixed, while $\beta$ increases in steps of 0.1. Figure~\ref{fig:sample-size-performance} plots the success rate and error composition versus $\beta$.

% \begin{figure}[h]
%     \centering
%     \includegraphics[width=0.5\linewidth]{figures/Figure_10k/Change Sample Size_success_10k_05.png}
%     \caption{Caption}
%     \label{fig:enter-label}
% \end{figure}

% \textit{Finding.} Our results confirm that increasing the sampling fraction $\beta$ improves verification accuracy, with the success rate reaching 0.98 at $\beta = 0.3$, which exceeds our target of $1-\delta = 0.9$. This indicates that $\beta = 0.3$ is near-optimal in this experimental setting, allowing us to allocate 70\% of EPR pairs for teleportation while maintaining high verification confidence. For this specific configuration, the average actual fidelity for the $(1-\beta)\mathcal{C}$ remaining pairs is measured around 0.97, which is above the threshold $1-\alpha=0.9$. For that reason, we do not have false positives in the test. In a general, both false negatives and negatives can occur.

\textit{Finding.} Increasing the sampling fraction $\beta$ improves verification accuracy, with success rate reaching 0.98 at $\beta=0.3$, above the $1-\delta=0.9$ target. Thus, $\beta=0.3$ is near-optimal, leaving 70\% of pairs for teleportation while maintaining high confidence. In this configuration, the mean fidelity of the $(1-\beta)\mathcal{C}$ remaining pairs is about 0.97, exceeding the $1-\alpha=0.9$ threshold. Hence no false positives occur, though both false negatives and positives are possible in general.

%Perhaps more significantly, the error decomposition reveals that the protocol exclusively produces false negatives with while no false positives is observed across all sampling fractions. This asymmetric error profile indicates that PEV operates conservatively, rejecting borderline cases that might compromise security. Such behavior is highly desirable in quantum security applications where false positives (accepting inadequate entanglement) could potentially enable information leakage, while false negatives (rejecting adequate entanglement) merely reduce efficiency without compromising security. {\color{red}{In general, false positive can happens, but with less frequency than false negative (see Figure \ref{} {\color{red}{???}})}
% }

\begin{figure}[h]
    \centering
    \includegraphics[width=0.6\linewidth]{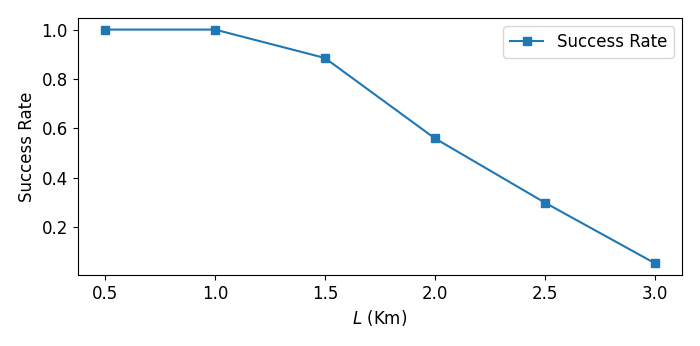}
    \includegraphics[width=0.6\linewidth]{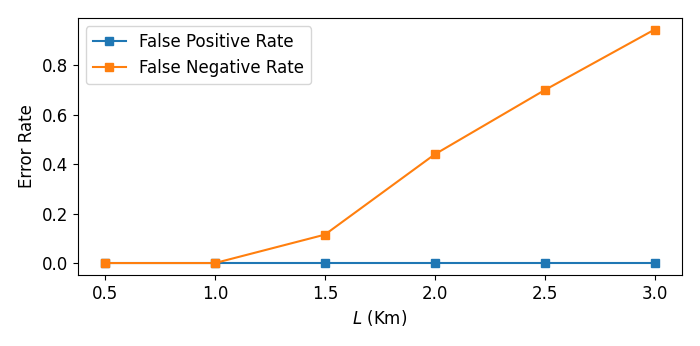}
    \includegraphics[width=0.6\linewidth]{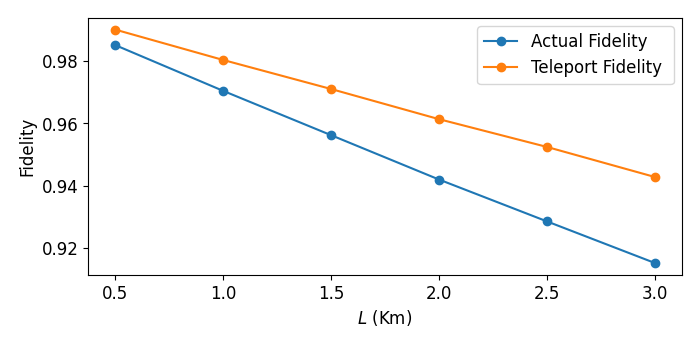}
    
    \caption{Top: PEV success rate vs. distance. Middle: Error decomposition into false positives/negatives. Bottom: Actual fidelity of remaining EPR pairs and post-teleportation fidelity. }
    \label{fig: distance effect}
\end{figure}

% \noindent
\textbf{Node Distance Effect.} 
We examine how node separation (varying from 0.5 to 3 km) affects verification performance. Figure \ref{fig: distance effect} shows how the success rate changes as a function of distance.

\textit{Finding.} For $L=1$ km, the success rate is 1. Increasing the distance to $L=1.5$ km reduces it to 89\%. In general, success declines with distance across all tested sample sizes, since longer channels degrade qubit quality and weaken CHSH verification. Using more test samples can partially mitigate this effect. As in previous experiments, the actual EPR fidelity remains above the threshold, so no false negatives occur. In case where the average fidelity is low, false positives will arise instead. We also simulated teleportation using accepted pairs: the fidelity of the teleported qubit slightly exceeds the actual entanglement fidelity, consistent with the data processing inequality (see Supplementary Material A).

% {\color{red}{We conjecture that there exists a critical point for $\beta$ at which a phase transition happens.}}

\begin{figure}[h]
    \centering
    \includegraphics[width=0.6\linewidth]{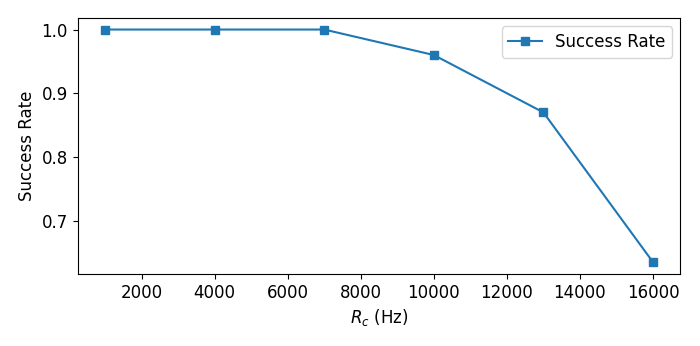}
    \includegraphics[width=0.6\linewidth]{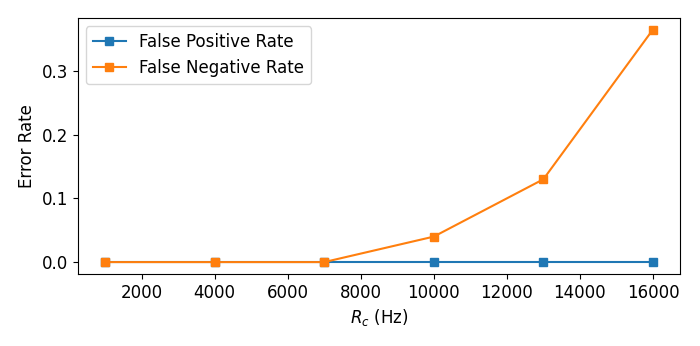}
    \includegraphics[width=0.6\linewidth]{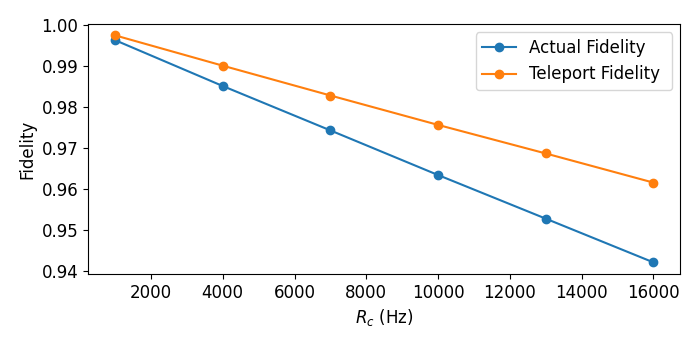}
    \caption{Top: PEV success rate vs. depolarization rate. Middle: Error decomposition into false positives/negatives. Bottom: Actual fidelity of remaining EPR pairs and post-teleportation fidelity.}
    \label{fig: depolarization effect}
\end{figure}

% \noindent
\textbf{Depolarization Rate Effect.}  
This experiment evaluates protocol performance under varying channel noise. The depolarization rate is swept from 1,000 to 16,000~Hz while all other parameters are fixed. The results are shown in Fig.~\ref{fig: depolarization effect}.

% \textit{Finding.} The success rate is maintained to be above 90\% when the depolarization rate $R_c$ is under $10,000$ Hz. We still obtain good performance with success rates above 80\% until $R_c=13,000$ Hz. As the channel gets noisier, the success rate drops, particularly at $R_c =16,000$ Hz, the value is below 0.7. This pattern is very similar to what we saw with increasing distance. Again, the direct solution for this problem is using more samples to achieve better 
% performance in noisy conditions. The teleportation fidelity upon acceptance is also investigated, and it behaves similarly to the previous setting's outcome. While the entanglement fidelity degrades with $R_c$, the post-teleportation fidelity of a single qubit state will be held a little bit above. Through our simulation framework, network designers can figure out which channel quality is necessary to reach a desired level of verification reliability if their resources and node distance remain unchanged.

\textit{Finding.} The success rate stays above 90\% for depolarization rates $R_c < 10{,}000$ Hz and remains above 80\% up to $R_c=13{,}000$ Hz, but drops below 0.7 at $R_c=16{,}000$ Hz. This pattern is very similar to what we saw with increasing distance. Teleportation fidelity upon acceptance shows the same pattern as before: while entanglement fidelity degrades with $R_c$, the fidelity of the teleported qubit remains slightly higher. Our framework thus helps network designers determine the channel quality needed to achieve reliable verification under fixed resources and node distances.

\begin{figure}[h]
    \centering
    \includegraphics[width=0.6\linewidth]{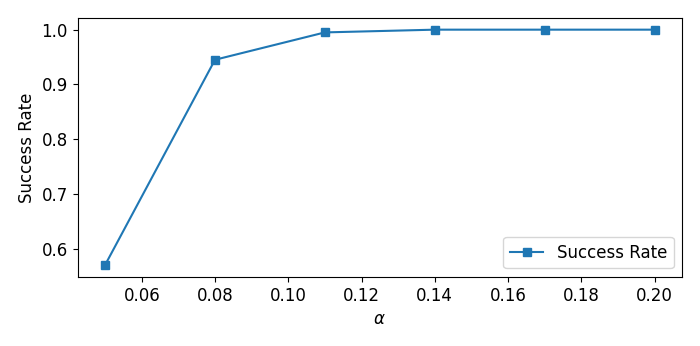}
    \caption{Success rate of PEV for different threshold values of $\alpha$.}
    \label{fig:alpha-effect}
\end{figure}

% \noindent
\textbf{PEV Performance with Varying Threshold.} 
We examine how the acceptance threshold $1-\alpha$ influences protocol accuracy. Intuitively, relaxing the criterion (larger $\alpha$) should increase the success rate. To test this, we vary $1-\alpha$ and measure the resulting performance in Fig.~\ref{fig:alpha-effect}.

% \textit{Finding.} Our results show that as $\alpha$ gets larger (meaning we accept lower quality entanglement), the protocol succeeds more often. For example, when we only require fidelity at least $0.8$ ($\alpha = 0.2$) instead of fidelity at least $0.95$ ($\alpha = 0.05$), the verification is more accurate. This creates an important trade-off for quantum networks. Stricter requirements (smaller $\alpha$) give better security but also push us to use more test samples, otherwise, verification can become less reliable. Easier requirements (larger $\alpha$) make verification more reliable but accept lower-quality entanglement that is less useful. An extra point worth noting is that the post-teleportation fidelity is always above the entanglement fidelity of the teleportation channel (as discussed in the two previous experimental results and also in Appendix A). This means one can consider a suitable fidelity threshold based on the desired post-teleportation fidelity.

\textit{Finding.} As $\alpha$ increases (i.e., lower fidelity is accepted), the protocol succeeds more often. For example, requiring fidelity $\geq 0.8$ ($\alpha=0.2$) yields higher accuracy than a stricter threshold of $0.95$ ($\alpha=0.05$). This highlights a key trade-off: small $\alpha$ provides stronger security but demands more samples to maintain reliability, while large $\alpha$ improves reliability but accepts weaker entanglement. Notably, post-teleportation fidelity is always above the underlying entanglement fidelity (see Supplementary Material A), so thresholds can be chosen based on the desired fidelity of teleported states.

The evaluation demonstrates protocol PEV’s performance across diverse operating conditions. With $C=10{,}000$ EPR pairs, our simulations identify regimes where the protocol is effective and where it becomes limited. For short distances ($L<1.5$ km), low noise ($R_c<8000$ Hz), and threshold $1-\alpha=0.9$, PEV achieves high success rates with reasonable sample sizes.  

The results emphasize key trade-offs for network designers: allocating pairs between testing and teleportation, balancing computational resources with node distance, and choosing between strict fidelity requirements and verification reliability. The optimal balance depends on application-specific security and efficiency needs.  

Our NetSquid framework enables designers to anticipate protocol performance under given conditions and tune parameters before deployment, reducing costly trial-and-error on physical quantum hardware.
\section{Conclusion}

We addressed the problem of efficiently verifying whether quantum connections between network nodes are strong enough for secure communication. Our approach connects two fundamental concepts—CHSH inequality violation and entanglement fidelity—and establishes bounds that reveal what one metric implies about the other. Building on this foundation, we proposed verification methods that make reliable decisions about entanglement quality with limited resources.  

Through realistic NetSquid simulations, we observed consistent patterns:  
\begin{itemize}
    \item Larger sample sizes improve verification accuracy.  
    \item Increasing distance or channel noise reduces success rates.  
    \item Stricter fidelity thresholds enhance security but lower efficiency, creating a trade-off with resource use.  
\end{itemize}

Beyond these findings, we introduced a modular simulation framework adaptable to different noise models, protocols, and resource conditions. This framework helps network designers evaluate trade-offs and optimize quantum resources, providing a cost-effective path toward practical deployment of entanglement verification in real quantum networks.

% % Acknowledgments (optional, before references)
% \section*{Acknowledgments}
% % ...

% --- References ---
% EITHER BibTeX:
\bibliographystyle{IEEEtran}
\bibliography{refs}
% \input{Appendix}
% % OR manual:
% \begin{thebibliography}{1}

% \end{thebibliography}

% --- Biographies (optional; after references) ---
% If including photos:
% \begin{IEEEbiography}[{\includegraphics[width=1in,height=1.25in,clip,keepaspectratio]{<photo>}}]{<Name>}
% Bio text...
% \end{IEEEbiography}

% If no photos:
% \begin{IEEEbiographynophoto}{<Name>}
% Bio text...
% \end{IEEEbiographynophoto}
\newpage
\section*{Appendix}

\section*{Appendix A: Details of proofs}\label{Appendix: Proofs}

\subsection*{A. Proof that $\mathcal{Q}$ is compact}

We prove that $\mathcal{Q}$, the set of density matrices, is both bounded and closed.

\begin{itemize}
\item \textit{Boundedness:}  
For any $\rho \in \mathcal{Q}$, we have $\mathrm{Tr}[\rho^2] \leq 1$.  
Thus all elements of $\mathcal{Q}$ lie within a fixed ball in the Hilbert–Schmidt norm, implying $\mathcal{Q}$ is bounded.

\item \textit{Closedness:}  
Let $\{\rho_n\} \subset \mathcal{Q}$ be a sequence converging to $\rho$.  
We show $\rho \in \mathcal{Q}$.

For any vector $\lvert \phi \rangle$:
\begin{align}
\langle \phi | \rho | \phi \rangle
&= \left\langle \phi \,\middle|\, \lim_{n \to \infty} \rho_n \,\middle|\, \phi \right\rangle
= \lim_{n \to \infty} \langle \phi | \rho_n | \phi \rangle.
\end{align}
The sequence $a_n = \langle \phi | \rho_n | \phi \rangle$ is bounded in the closed set $\mathbb{R}^+ \cup \{0\}$, hence its limit  
$\langle \phi | \rho | \phi \rangle \in \mathbb{R}^+ \cup \{0\}$.  
Therefore $\rho$ is positive semidefinite.  
Moreover, $\mathrm{Tr}[\rho] = \lim_{n \to \infty} \mathrm{Tr}[\rho_n] = 1$.  
Thus $\rho \in \mathcal{Q}$, proving $\mathcal{Q}$ is closed.  
Since $\mathcal{Q}$ is bounded and closed in a finite-dimensional space, it is compact. \qed
\end{itemize}

\subsection*{B. Error probability of Protocol EV}

The probability of error is
\begin{align}
\mathbb{P}[\mathrm{Error}]
&= \mathbb{P}[\bar{S} \geq \theta_{\mathrm{th}},\, F \leq 1 - 3\alpha]
 + \mathbb{P}[\bar{S} < \theta_{\mathrm{th}},\, F \geq 1 - \alpha] \notag\\
&\leq \mathbb{P}[\bar{S} \geq \theta_{\mathrm{th}},\, S \leq 2\sqrt{2} - 6\sqrt{2}\,\alpha] \notag\\
&\quad+ \mathbb{P}[\bar{S} < \theta_{\mathrm{th}},\, S \geq 2\sqrt{2} - 4\sqrt{2}\,\alpha] \notag\\
&\leq \mathbb{P}[\bar{S} - S \geq \sqrt{2}\,\alpha,\, S \leq 2\sqrt{2} - 6\sqrt{2}\,\alpha] \notag\\
&\quad+ \mathbb{P}[\bar{S} - S < -\sqrt{2}\,\alpha,\, S \geq 2\sqrt{2} - 4\sqrt{2}\,\alpha] \notag\\
&\leq \mathbb{P}[|\bar{S} - S| \geq \sqrt{2}\,\alpha] + \mathbb{P}[|\bar{S} - S| > \sqrt{2}\,\alpha] \notag\\
&= 2\,\mathbb{P}[|\bar{S} - S| \geq \sqrt{2}\,\alpha].
\end{align}

\subsection*{C. Post-teleportation fidelity vs.\ entanglement fidelity}

Let $\mathcal{E}$ be the CPTP map describing the teleportation protocol, which transforms the joint system  
(entangled resource pair + input qubit) into the teleported output.  
Let $\rho$ be the state of the entangled pair, with entanglement fidelity $F(\rho)$, and let $\sigma$ be an arbitrary input state.

By the data processing inequality for fidelity:
\begin{equation}
F\!\left(\rho \otimes \sigma,\, \ket{\Phi^+}\!\bra{\Phi^+} \otimes \sigma\right)
\leq F\!\left(\mathcal{E}(\rho \otimes \sigma),\, \mathcal{E}(\ket{\Phi^+}\!\bra{\Phi^+} \otimes \sigma)\right).
\end{equation}
The left-hand side equals the entanglement fidelity $F(\rho)$.  
Since $\mathcal{E}(\ket{\Phi^+}\!\bra{\Phi^+} \otimes \sigma) = \sigma$ (perfect teleportation),  
the right-hand side is the fidelity between the teleported state and the input $\sigma$, i.e., the post-teleportation fidelity.  
Hence the post-teleportation fidelity is at least the entanglement fidelity. \qed

\section*{Appendix B: Decision procedure based on DFE}

\textit{EPR (Bell) state fidelity estimation.}\label{Appendix: DFE}
For the target state $|\Phi^+\rangle$, the only nonzero Pauli overlaps are
\begin{align*}
\langle XX\rangle_{|\Phi^+\rangle} &= +1,\quad\quad \langle YY\rangle_{|\Phi^+\rangle} = -1\\
\langle ZZ\rangle_{|\Phi^+\rangle} &= +1,\quad\quad \langle II\rangle_{|\Phi^+\rangle} = +1,
\end{align*}with $p_k = 1/4$ for $W_k \in \{II,XX,-YY,ZZ\}$ and $0$ otherwise. Since $|\Phi^+\rangle$ is a stabilizer state, the estimation complexity is dimension-independent. The DFE fidelity estimator simplifies to an average of \emph{signed} two-qubit correlators:
\[
\hat{F} = \frac14 \left(
1 + {\langle XX\rangle}_\rho - {\langle YY\rangle}_\rho
+ {\langle ZZ\rangle}_\rho \right),
\]
where the $II$ term is known exactly and requires no measurement shots.  

To test $H_0: F \le a$ versus $H_1: F \ge b$ with $b>a$ and equal risks $\alpha = \beta = \delta$, use the midpoint threshold $t = (a+b)/2$ and decide $H_1$ if $\hat{F} \ge t$, else $H_0$. Hoeffding’s inequality for bounded i.i.d.\ samples gives the number of settings
\[
\ell = \frac{2}{(b-a)^2} \ln\frac{1}{\delta},
\]
and choosing per-setting accuracy $\varepsilon_{\mathrm{meas}} \le (b-a)/4$ leads to a total copy complexity
\[
\mathbb{E}[M] = \frac{32}{(b-a)^2} \ln\frac{2}{\delta_{\mathrm{meas}}},
\]
which is independent of the Hilbert space dimension. In practice, one may sample uniformly from $\{XX,YY,ZZ\}$ and add the $II$ term deterministically, reweighting the estimator to remain unbiased.

\section*{Appendix C: Noise Models and Bound Computation}\label{Appendix: Noise models and GHZ}

In our simulations, we consider several noise channels acting on the initial maximally entangled Bell state 
$\lvert \Phi^+ \rangle = (\lvert 00 \rangle + \lvert 11 \rangle)/\sqrt{2}$. 
These channels capture both Pauli-type and non-Pauli decoherence processes relevant in optical, atomic, and solid-state entanglement distribution experiments.

For each noise model, we generate the noisy state $\rho$ and compute:
\begin{itemize}
    \item \textbf{True fidelity:} $F_{\Phi^+} = \langle \Phi^+ | \rho | \Phi^+ \rangle$.
    \item \textbf{CHSH bounds:} The analytical lower and upper bounds from our CHSH-based protocol.
    \item \textbf{GHZ($n{=}2$) bound:} The analytic formula 
    $L_{\mathrm{GHZ2}} = \tfrac12\left(\langle XX\rangle_\rho - \langle YY \rangle_\rho\right)$ obtained from equatorial-basis correlation statistics.
\end{itemize}

\subsection{Noise Channels}

\textit{1. Werner State-Depolarizing Channel}  
Mixes the maximally entangled state with the maximally mixed state:
\begin{equation}
\rho_{\text{Werner}}(p) = p \lvert \Phi^+ \rangle \langle \Phi^+ \rvert + (1-p) \frac{\mathbb{I}_4}{4},
\end{equation}
where $p \in [0,1]$ is the visibility.  
Models isotropic noise that reduces all correlations equally, yielding a Bell-diagonal output.  
See Fig. 3a in the main text.

\textit{2. Amplitude Damping (Non-Pauli)}  
Models energy relaxation from $\lvert 1\rangle$ to $\lvert 0\rangle$:
\begin{align}
K_0 &= \begin{pmatrix} 1 & 0 \\ 0 & \sqrt{1-\gamma} \end{pmatrix}, &
K_1 &= \begin{pmatrix} 0 & \sqrt{\gamma} \\ 0 & 0 \end{pmatrix},
\end{align}
where $\gamma \in [0,1]$ is the damping probability.  
Relevant to systems with spontaneous emission or photon loss.  
Results in Fig. 3b.

\textit{3. Dephasing (Pauli-$Z$ Channel)}  
This channel models the loss of phase coherence without affecting populations.  
It is described as a Pauli channel with only $Z$ flips:
\begin{equation}
\mathcal{E}_Z(\rho) = (1-q_Z)\rho + q_Z\,Z\rho Z,
\end{equation}
with $q_Z \in [0,1]$ the dephasing probability.  
In the computational basis, the diagonal elements remain unchanged while the off-diagonal terms are suppressed by a factor $(1-2q_Z)$:
\[
\rho_{01} \mapsto (1-2q_Z)\,\rho_{01}.
\]

This map is mathematically equivalent to the commonly used \emph{phase damping channel}, defined by Kraus operators
\[
K_0=\sqrt{1-\gamma}\,I,\quad 
K_1=\sqrt{\gamma}\,|0\rangle\langle 0|,\quad 
K_2=\sqrt{\gamma}\,|1\rangle\langle 1|,
\]
under the parameter mapping $\gamma = 2q_Z$.  
The two descriptions differ mainly in interpretation: phase damping arises from microscopic models of decoherence (e.g., photon scattering), while Pauli-$Z$ dephasing is the abstract form for qubit noise model.

\medskip
These models span uncorrelated local noise, pure dephasing, and relaxation.  
The Pauli channels produce Bell-diagonal states where both CHSH and GHZ2 bounds are analytic; the non-Pauli channels require full density-matrix evolution for bound evaluation.

% \subsection{Why amplitude damping and phase noise affect GHZ verification more than CHSH}
% % \begin{algorithm}
% \begin{algorithm}[h!]
% \caption{GHZ Protocol for EPR State Verification}
% \begin{algorithmic}[1]
% \STATE \textbf{Input:} Integer $N$, where the total number of measured copies of $\rho$ is $N$.
% \STATE Alice and Bob prepare $N$ copies of $\rho$, then perform the following experiment:
% \FOR{each copy $k \in \{1,2,...,N\}$}
%     \STATE Randomly choose measurement angles $\theta_A, \theta_B$ with constraint $\frac{\theta_A + \theta_B}{\pi} \in \mathbb{N}$.
%     \STATE Alice measures in basis $\{\frac{\ket{0} + e^{i\theta_A}\ket{1}}{\sqrt{2}}, \frac{\ket{0} - e^{i\theta_A}\ket{1}}{\sqrt{2}}\}$.
%     \STATE Bob measures in basis $\{\frac{\ket{0} + e^{i\theta_B}\ket{1}}{\sqrt{2}}, \frac{\ket{0} - e^{i\theta_B}\ket{1}}{\sqrt{2}}\}$.
%     \STATE Record outcomes $(Y_A^k, Y_B^k) \in \{0,1\} \times \{0,1\}$.
% \ENDFOR
% \STATE \textbf{Computation:} Calculate success rate:
% \[
% \hat{P} = \frac{1}{N} \sum_{k=1}^{N} \mathbb{1}\left[Y_A^k + Y_B^k \equiv \frac{\theta_A^k + \theta_B^k}{\pi} \pmod{2}\right].
% \]
% \STATE \textbf{Output:} The estimated fidelity lower bound $2\hat{P} - 1$.
% \end{algorithmic}
% \label{Alg: GHZ-Protocol}
% \end{algorithm}

\subsection{Why amplitude damping and phase noise affect GHZ verification more than CHSH}

\begin{algorithm}[t]
\caption{GHZ($n{=}2$) Protocol for EPR State Verification}
\begin{algorithmic}[1]
\STATE \textbf{Input:} Integer $N$ (number of measured copies of $\rho$).
\STATE Prepare $N$ copies of $\rho$.
\FOR{$k=1$ to $N$}
    \STATE Pick $\theta_A^{(k)},\theta_B^{(k)}$ uniformly at random such that $(\theta_A^{(k)}+\theta_B^{(k)})/\pi \in \{0,1\}$.
    \STATE Alice measures 
    \[
        M_A(\theta_A^{(k)})=\cos\theta_A^{(k)}\, X+\sin\theta_A^{(k)}\, Y,
    \]
    obtaining $a_k\in\{\pm1\}$.
    \STATE Bob measures 
    \[
        M_B(\theta_B^{(k)})=\cos\theta_B^{(k)}\, X+\sin\theta_B^{(k)}\, Y,
    \]
    obtaining $b_k\in\{\pm1\}$.
    \STATE Map to bits: $y_A^{(k)}=\mathbf{1}[a_k=-1]$, $y_B^{(k)}=\mathbf{1}[b_k=-1]$.
\ENDFOR
\STATE Compute
\[
\widehat{P}=\frac{1}{N}\sum_{k=1}^N \mathbf{1}\!\left[
y_A^{(k)} \oplus y_B^{(k)} \equiv \frac{\theta_A^{(k)}+\theta_B^{(k)}}{\pi} \;(\mathrm{mod}\;2)
\right].
\]
\STATE \textbf{Output:} $L_{\mathrm{GHZ2}} = 2\widehat{P}-1$.
\end{algorithmic}
\label{Alg:GHZ2-Protocol}
\end{algorithm}

\subsubsection*{1. Correlation observable for GHZ($n{=}2$)}
For any angles $\theta_A,\theta_B$, define the equatorial observables
\begin{align}
M_A(\theta_A) &= \cos\theta_A\,X+\sin\theta_A\,Y, \\
M_B(\theta_B) &= \cos\theta_B\,X+\sin\theta_B\,Y,
\end{align}
whose eigenbases are 
\(\big\{\tfrac{|0\rangle \pm e^{i\theta_A}|1\rangle}{\sqrt{2}}\big\}\) and similarly for $M_B$.  
McCutcheon \textit{et al.}~\cite{McCutcheon2016-GHZ-verification} originally stated the protocol in terms of measurement bases;  
here we rewrite it in the equivalent observable form (Algorithm~\ref{Alg:GHZ2-Protocol}) and map outcomes $\pm 1$ to bits $0$ and $1$.

The correlation for a state $\rho$ is
\begin{align}
E(\theta_A,\theta_B) &= \mathrm{Tr}\!\left[\rho\, M_A(\theta_A)\!\otimes\! M_B(\theta_B)\right] \notag\\
&= \cos\theta_A\cos\theta_B\,\langle XX\rangle \notag\\
&\quad+ \cos\theta_A\sin\theta_B\,\langle XY\rangle \notag\\
&\quad+ \sin\theta_A\cos\theta_B\,\langle YX\rangle \notag\\
&\quad+ \sin\theta_A\sin\theta_B\,\langle YY\rangle.
\label{eq:equatorial-corr}
\end{align}

\subsubsection*{2. Success probability}
Let $k\in\{0,1\}$ denote $(\theta_A+\theta_B)/\pi \bmod 2$.  
With the mapping $+1\mapsto 0$, $-1\mapsto 1$, the parity test  
$y_A \oplus y_B \equiv k$ is equivalent to $a\,b = (-1)^k$.  
The conditional success probability is
\begin{equation}
P_{\mathrm{succ}}(\theta_A,\theta_B)
= \frac{1+(-1)^k\,E(\theta_A,\theta_B)}{2}.
\label{eq:succ-prob}
\end{equation}

\subsubsection*{3. Averaging over allowed angles}
Angles are chosen uniformly with $\theta_A+\theta_B=k\pi$. Two cases:

\begin{itemize}
    \item If $k=0$, set $\theta_B=-\theta$, then $\cos\theta_B=\cos\theta$, $\sin\theta_B=-\sin\theta$.  
From \eqref{eq:equatorial-corr}, averaging over $\theta\sim\mathrm{Unif}[0,2\pi)$ gives
\begin{equation}
\mathbb{E}[E(\theta,-\theta)] 
= \tfrac12\big(\langle XX\rangle - \langle YY\rangle\big).
\label{eq:Eavg-k0}
\end{equation}

\item If $k=1$, set $\theta_B=\pi-\theta$, then $\cos\theta_B=-\cos\theta$, $\sin\theta_B=\sin\theta$.  
Similarly,
\begin{equation}
\mathbb{E}[E(\theta,\pi-\theta)] 
= -\tfrac12\big(\langle XX\rangle - \langle YY\rangle\big).
\label{eq:Eavg-k1}
\end{equation}
\end{itemize}
In \eqref{eq:succ-prob}, the factor $(-1)^k=-1$ cancels the minus sign, giving the same average as $k=0$.

\subsubsection*{4. Final expression}
Since both $k$ values occur with equal probability:
\begin{align}
\mathbb{E}[\widehat{P}]
&= \frac12 + \frac{\langle XX\rangle - \langle YY\rangle}{4}, \\
\Rightarrow\quad
\mathbb{E}[\,2\widehat{P}-1\,]
&= \frac{\langle XX\rangle - \langle YY\rangle}{2}.
\label{eq:GHZ2-LB}
\end{align}
Thus $L_{\mathrm{GHZ2}}$ is an unbiased estimator of the analytic lower bound
\begin{equation}
L_{\mathrm{GHZ2}}^{\mathrm{(analytic)}} 
= \frac12\big(\langle XX\rangle - \langle YY\rangle\big).
\end{equation}

\end{document}